%% file: ms.tex
\documentclass[useAMS,usenatbib]{mn2e}

\input psfig.sty
\usepackage{graphicx}
\usepackage{xspace}

\input{macros.tex}

\input{addresses.tex}
\newcommand\nodata{ ~$\cdots$~ }%

\title[SWELLS III. Disfavouring heavy IMFs for spirals] {The SWELLS
survey. III. Disfavouring ``heavy'' initial mass functions for spiral lens galaxies}
    
\author[Brewer \etal]{%
  Brendon~J.~Brewer$^{1}$\thanks{\breweremail}, 
  Aaron~A.~Dutton$^{1,2,3}$\thanks{\cita},
  Tommaso~Treu$^1$\thanks{\packard}, 
\newauthor{%
  Matthew~W.~Auger$^{1,4}$,
  Philip~J.~Marshall$^{5,6}$,
  Matteo~Barnab\`e$^{5}$,
  }
\newauthor{%
  Adam~S.~Bolton$^7$, 
  David~C.~Koo$^3$,
  L\'eon~V.~E.~Koopmans$^8$}
  \medskip\\
  $^1$\ucsb\\
  $^2$\uvic\\
  $^3$\lick\\
  $^4$\cambridge\\
  $^5$\kipac\\
  $^6$\oxford\\
  $^7$\utah\\
  $^8$\kapteyn\\
}

\begin{document}
             
\date{To be submitted to MNRAS}
             
\pagerange{\pageref{firstpage}--\pageref{lastpage}}\pubyear{2012}

\maketitle           

\label{firstpage}


\begin{abstract}
  We present gravitational lens models for 20 strong gravitational
  lens systems observed as part of the Sloan WFC Edge-on Late-type
  Lens Survey (SWELLS) project. Fifteen of the lenses are taken from
  paper I while five are newly discovered systems.  The systems are
  galaxy-galaxy lenses where the foreground deflector has an inclined
  disc, with a wide range of morphological types, from late-type
  spiral to lenticular.  For each system, we compare the total mass
  inside the critical curve inferred from gravitational lens modelling
  to the stellar mass inferred from stellar population synthesis (SPS)
  models, computing the stellar mass fraction $f^* \equiv M_{\rm
  SPS}/M_{\rm lens}$.  We find that, for the lower mass SWELLS
  systems, adoption of a Salpeter stellar initial mass function (IMF)
  leads to estimates of $f^*$ that exceed 1. This is unphysical, and
  provides strong evidence against the Salpeter IMF being valid for
  these systems. Taking the lower mass end of the SWELLS sample
  ($\sigma_{\rm SIE} < 230 \kms$), we find that the IMF is lighter (in
  terms of stellar mass-to-light ratio) than Salpeter with 98\%
  probability, and consistent with the Chabrier IMF and IMFs between the two. This result is
  consistent with previous studies of spiral galaxies based on
  independent techniques. In combination with recent studies of
  massive early-type galaxies that have favoured a heavier
  Salpeter-like IMF, this result strengthens the evidence against a
  universal stellar IMF.
\end{abstract}



\section{Introduction}

The stellar initial mass function (IMF) is a fundamental property of a
simple stellar population, and plays a key role in many astrophysical
problems. For instance, the determination of galaxy stellar masses and
star formation rates are uncertain at the factor of $\sim 2$ level due
to uncertainty in the IMF \citep[e.g.][]{2009ApJ...699..486C,
  2007MNRAS.378.1550P}.

Observations of the solar neighbourhood favor an IMF with a power law
slope at high masses \citep{1955ApJ...121..161S}, and a turnover at
low masses \citep{2001MNRAS.322..231K, 2003PASP..115..763C}. While
other spiral galaxies have IMFs consistent with that of the Milky Way
\citep{2001ApJ...550..212B, 2011ApJ...739L..47B, 2011MNRAS.416..322D,
2011MNRAS.417.1621D, 2011arXiv1110.2536S}, there is increasing
evidence from a variety of observations that the IMF is
non-universal. Specifically, massive early-type galaxies at
low redshifts require IMFs that are heavier than in the Milky Way
 \citep[e.g.][]{2011MNRAS.417.3000S, 2011arXiv1111.4215S, 2010ApJ...721L.163A}, 
and possibly heavier than Salpeter in the most massive ellipticals
\citep{2010Natur.468..940V}. Other lensing and
dynamical studies of early-type galaxies also suggest that the IMF
may be heavy \citep{2010ApJ...709.1195T, 2011MNRAS.415..545T}
although these observations may also be explained by a trend in the properties of
the dark matter haloes. Recently, variations in the IMF
of early-type galaxies have also
been reported by \citet{2012arXiv1202.3308C} and \citet{2011arXiv1111.2905D}.

One of the simplest ways of constraining the galaxy averaged IMF is
through comparing total masses (within some aperture), as derived by
kinematics and/or strong gravitational lensing and referred to as
gravitational masses henceforth, with stellar masses derived from
stellar population synthesis (SPS) models
\citep{2006MNRAS.366.1126C,2010MNRAS.409L..30F, 2011MNRAS.415..545T}. While this method
typically only provides strong upper limits to the stellar mass due to
the unknown dark matter fraction, as well as the unknown mass in gas,
dust etc., the upper limits are robust.

Dark matter fractions in galaxies are expected to increase with
radius, so the strongest constraints on the IMF (from total mass
measurements) come from measuring mass-to-light ratios at small
galactic radii. However, at small radii kinematics are harder to
interpret, as there is usually a mix of rotation and dispersion. This
is where strong gravitational lensing is particularly useful and
unique, because it measures projected mass independent of the
dynamical state of the deflector.

Assuming a universal IMF, the dark matter fractions within the
effective radius of early-type galaxies decrease with decreasing mass
(down to a stellar mass of $\simeq 10^{10}\Msun$)
\citep[e.g.][]{2004NewA....9..329P, 2006MNRAS.370.1106G,
  2010ApJ...724..511A, 2011MNRAS.416..322D, 2011MNRAS.415.2215B}. Thus
lower mass elliptical galaxies are expected to place stronger
constraints on the IMF than higher mass elliptical galaxies
\citep[although high mass systems can place strong constraints as well
through dynamics, e.g.][]{2011MNRAS.417.3000S}.  If the bulges of
spiral galaxies follow the same scaling relations as elliptical
galaxies then we would expect that bulges of mass $\simeq 10^{10}\Msun$ --
where the dark matter fraction is minimised -- would provide the
strongest constraints on the IMF if the IMF is universal
\cite[e.g.][]{2010ARA&A..48..339B,2011arXiv1112.3340K}. Conversely, if
the IMF is not universal, measuring it as a function of galaxy mass,
morphology, and environment should help understand the physical
mechanisms that determine it.

Measuring the absolute normalisation of stellar IMF as a function of
galaxy type and mass is one of the main goals of the Sloan WFC Edge-on
Late-type Lens Survey (SWELLS; Treu \etal 2011, Dutton \etal 2011b;
hereafter papers I and II, respectively).

In this paper, we present Hubble Space Telescope ({\it HST}) multicolour
imaging of the complete SWELLS sample, including new data from
cycle-18 observations, and use strong lensing models to constrain the
stellar initial mass function of spiral galaxies. In particular we use
precision measurements of the total mass within the critical curve and
therefore place an upper limit on the amount of stellar mass. We
combine this inferences with stellar masses derived from stellar
population synthesis models to obtain strong upper limits on the
normalisation of the stellar IMF.  

The galaxies in the SWELLS sample have total stellar masses (assuming
a Chabrier IMF) of $M_*\simeq 4-30 \times 10^{10}\Msun$, and bulge
masses of $\simeq 2-10\times 10^{10}\Msun$
\citep{2011MNRAS.417.1601T}. In order to explore the variation of the
IMF with stellar mass, we combine our sample with the more more
massive elliptical galaxy lenses from the SLACS survey
\citep{2009ApJ...705.1099A} with $\Mstar \simeq 10-50\times
10^{10}\Msun$.

Throughout this paper, we assume a flat $\Lambda$CDM cosmology with
present day matter density, $\Omega_{\rm m}=0.3$, and Hubble
parameter, $H_0=70 \rm\,km\,s^{-1}\,Mpc^{-1}$.

\section{The Data}
\label{sec:data}

In this section we briefly summarise the sample selection procedure
(\S~\ref{ssec:sel}; more details can be found in paper I) and then
present the new cycle-18 {\it HST} observations in Section~\ref{ssec:hst}.

\subsection{Sample Selection}
\label{ssec:sel}

The sample of spiral lens galaxy candidates was selected from the SDSS
database as discussed in paper I. Briefly, the spectroscopic database
was searched for composite spectra, consisting of foreground spectrum
superimposed with multiple emission lines at a higher redshifts,
following the procedures developed for the SLACS \citep[][hereafter
B06 and B08]{2006ApJ...638..703B,2008ApJ...682..964B} and BELLS
\citep{2012ApJ...744...41B} surveys.  In addition, however, in order
to identify inclined late-type galaxies suitable for rotation curve
measurements, the deflector galaxies were selected to have axis ratios
$q = b/a \lta 0.6$ with a preference for galaxies with edge-on discs
and emission lines indicating star formation. A priori lensing
probabilities were estimated based on the SDSS stellar velocity
dispersion and stellar Einstein Radius (paper I).

\subsection{HST observations and surface photometry} 
\label{ssec:hst}

\begin{figure*}
\begin{center}
\includegraphics[clip,scale=0.75]{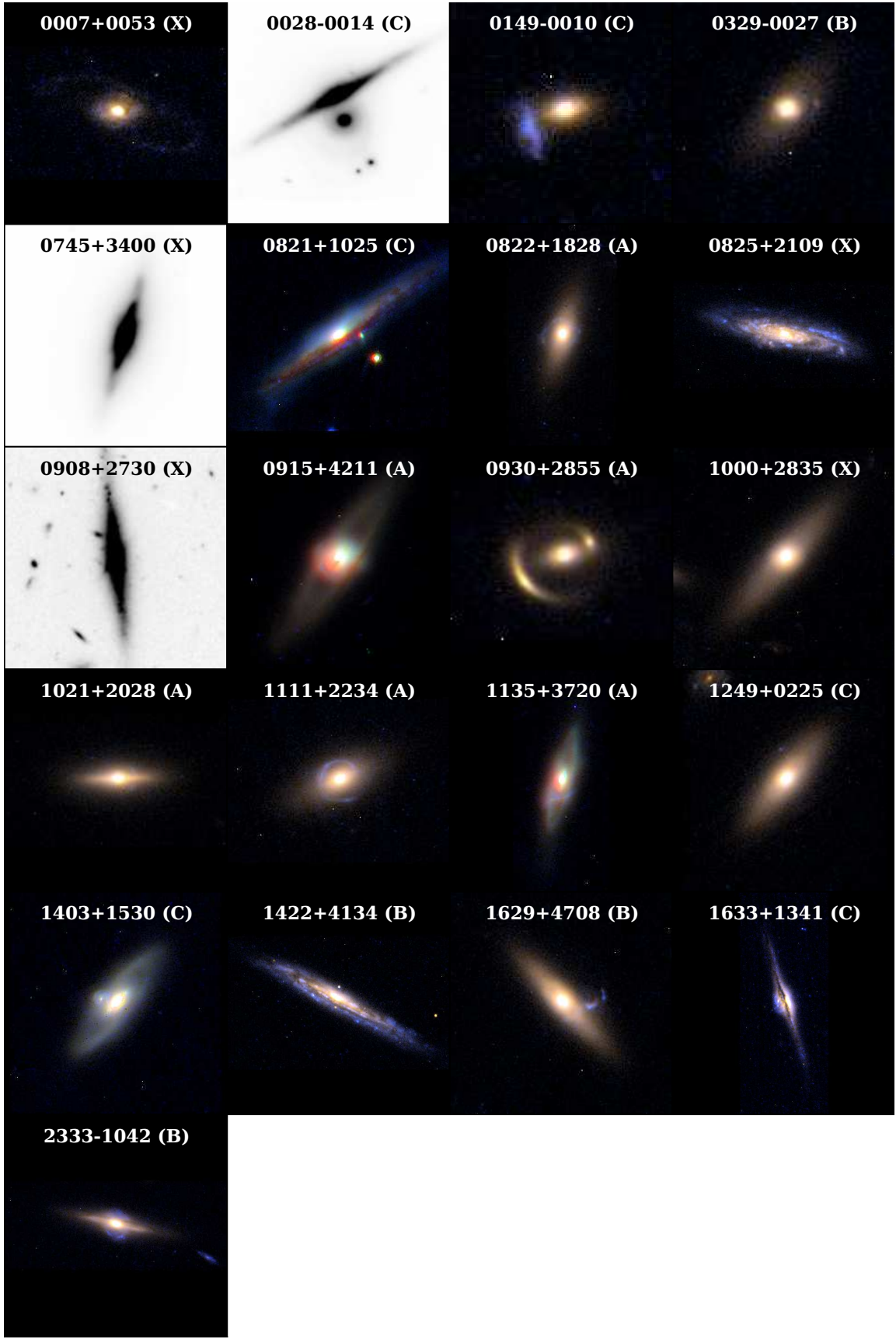}
\end{center}
\caption{Images of the 21 candidate lens systems oberved with the
{\it HST} during Cycle 18; the grade of each system is provided in
parentheses next to the name.\label{figure_cycle18_montage}}
\end{figure*}

The highest probability lens candidates were followed up with {\it HST} in
supplemental cycle 16s program (GO-11978; PI: Treu) and are presented
in paper I. In addition, during the same program, multiband imaging
was obtained for disc lens galaxies identified as part of the SLACS
survey (GOs 10174, 10494, 10798, 11202 PI: Koopmans; GO 10587, 10886,
PI: Bolton).  Lower probability systems were included in a cycle 18
program (GO-12292; PI: Treu), aiming to extend the sample to later type
spirals, and are presented here for the first time. Two candidates
confirmed from the ground with adaptive optics (paper I) were also
included in the cycle 18 program to collect optical multiband
photometry. Twenty one candidates were observed in cycle-18. Eighteen
of those were observed with the refurbished Advanced Camera for
Surveys (ACS) for one orbit each, split equally between filters F435W
and F814W. Two dithered exposures were obtained for each filter to aid
in cosmic ray and defect removal and to help improve the sampling of the
PSF. Four candidates, including one observed with ACS were observed
with the infrared channel of the Wide Field Camera 3 for a full orbit
through filter F160W. Each orbit was split into four dithered
exposures to aid in cosmic ray and defect removal and help improve the
sampling of the PSF. The {\it HST} images were reduced using standard
multidrizzle techniques. As expected the fraction of grade 'A' and 'B'
lenses was lower for cycle 18 than for previous cycles, 6/21 and 10/21
respectively; images for these 21 systems are displayed in Figure
\ref{figure_cycle18_montage}.

\begin{figure*}
\begin{center}
\includegraphics[clip,scale=0.85]{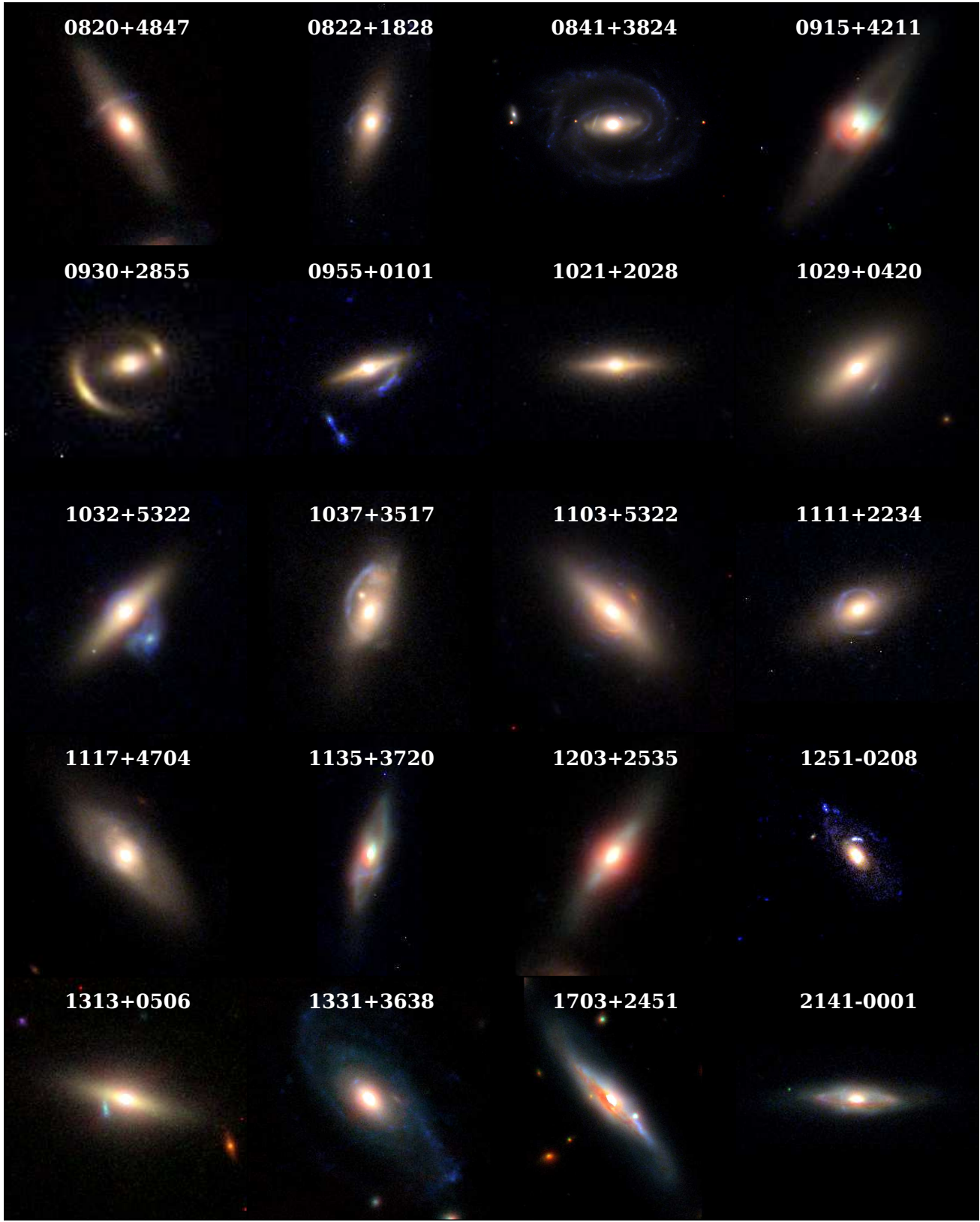}
\end{center}
\caption{Images of the 20 grade `A' lens systems that comprise the
SWELLS sample. \label{figure_grade_A_montage}}
\end{figure*}

In total, after {\it HST} imaging, the SWELLS sample consists of 20 secure
lenses (grade 'A'), 6 probable lenses (grade 'B'), and 12 possible
lenses (grade 'C'). Grade 'A' lenses are defined as having unambiguous
multiple images successfully reproduced by a reasonably simple lens
model. Grade 'B' lenses are defined as having evidence for strong
lensing, even though the quality of the data is not sufficient to
identify unambiguously multiple images and/or construct a lens
model. This definition is quite conservative and we expect that with
deeper data or narrow band imaging data to facilitate removal of the
foreground deflector light most 'B' grade systems would be confirmed
as lenses. Grade 'C' are possible lenses where we expect that even
with significantly better data only a minority of the systems would be
confirmed as lenses. A colour montage of the 20 grade 'A' lenses from
SWELLS is shown in Figure \ref{figure_grade_A_montage}.

For this paper we focus our analysis on the grade 'A' lenses, plus one
grade 'B' lens (J1422+4134), for which we were able to obtain a
plausible lens model. However, in light of the uncertain
identification of this one system, we will use only the grade 'A'
lenses for our inference on the stellar initial mass function. We also
excluded one grade 'A' lens, J1037+3517, since the morphology of the
lens is complex (involving a merger) and the modelling approach of
this paper was unlikely to be successful for this system.  A summary
of the basic properties of the SWELLS lenses together with the
inferred parameters of the gravitational lens models is given in
Table~\ref{tab:sample}.  A summary of the basic properties of the
SWELLS {\it HST} Cycle 18 targets is given in Table~\ref{tab:sample2}.

Multiband surface photometry was derived as described in paper I, by
fitting two-component (bulge+disc) models using \SPASMOID
\citep{2011ApJ...726...59B}.  The model consists of an $n=4$ S\'ersic
profile \citep[i.e. a de Vaucouleurs profile, ][]{Sersic1968, devauc} for the bulge and an $n=1$
exponential disc component. Estimates of the model parameters, including stellar masses and mass-to-light ratios, are
given in~Tables~\ref{tab:sample3},~\ref{tab:sample4} and~\ref{tab:sample5}.

\begin{table*}
\begin{center}
\scriptsize\begin{tabular}{cccccccccccccc}
\hline
\hline 
ID    & RA   & DEC  & $\zd$ & $\zs$& $b$ & $\sigma_{\rm SIE}$ & $q$   
& $M_{\rm lens}$ & $f^*_{\rm Chab}$ & $f^*_{\rm Salp}$ & Ref & Band \\
      &      &      &       &      &    ('')     &   ($\kms$)   &         &   ($10^{10} \Msun$)     \\
(1)   & (2)  & (3)  & (4)   & (5)  & (6)   & (7)        & (8)  & (9) & (10) & (11) & (12) & (13)\\
\hline
\input{catalogs/simple_table1.tex}

\hline \hline
\end{tabular}
\caption{Summary of basic properties of modelled grade 'A' lenses
  (plus the modelled grade 'B' lens J1422+4134) from the SWELLS sample
  of spiral lens galaxies.  Col.~1 lists the lens ID; Cols.~2 and~3
  the coordinates (J2000 in degrees); Cols~4 and~5 give deflector and
  source redshifts; Col.~6 the Einstein Radius of the lens model;
  Col.~7 the velocity dispersion of the Singular Isothermal Ellipsoid plus external shear
  (SIE+$\gamma$) lens model; Col~8. the axis ratio of the lens model; Col.~9
  the projected mass within the critical curve of the SIE lens model;
  Cols.~10 \& 11 the ratio between the stellar mass and lensing mass
  within the critical curve, assuming Chabrier and Salpeter IMF,
  respectively; Col.~12 is the reference for the lens discovery
  (1=B06, 2=B08, 3=\citet{2011MNRAS.417.1601T}, 4=this paper); Col.~13
  is the band used for the lens modelling. \label{tab:sample}}
\end{center}
\end{table*}


\begin{table*}
\begin{center}
  \scriptsize\begin{tabular}{ccccccccccccc} \hline \hline
    ID    & RA   & DEC  & $\zd$ & $\zs$& Grade & $q_{\rm SDSS}$ & Ref & Filter(s) \\
           (1)   & (2)  & (3)  & (4)   & (5)  & (6) & (7) & (8) & (9) \\
    \hline \input{catalogs/simple_table2.tex} \hline \hline
\end{tabular}
\caption{Summary of basic properties of the SWELLS {\it HST} Cycle 18 targets. 
Col.~1 lists the lens ID; Cols.~2 and 3 the coordinates (J2000 in degrees); 
Cols.~4 and 5 give deflector and source redshifts; 
Col.~6 the lens grade (A=secure, B=probable; C=possible, X=not a lens);  Col.~7 the axis ratio as measured by SDSS;
Col.~8 the reference for the discovery of the lens 
(3=\citet{2011MNRAS.417.1601T}, 4=this paper); Col.~9 the {\it HST} filters. 
\label{tab:sample2}}
\end{center}
\end{table*}


\begin{table*}
\begin{center}
  \scriptsize\begin{tabular}{ccccccccccccc} \hline \hline
    ID    & $B_{\rm bulge}$ & $I_{\rm bulge}$ & $H_{\rm bulge}$ & $K_{\rm bulge}$ & $B_{\rm disc}$ & $I_{\rm disc}$ & $H_{\rm disc}$ & $K_{\rm disc}$ & $R_{\rm e,bulge}$ & $q_{\rm bulge}$ & $R_{\rm e,disc} $ & $q_{\rm disc}$ \\
        &   &   &   & & & & &   & [arcsec] &   & [arcsec] &\\
   (1)   & (2)  & (3)  & (4)   & (5)  & (6) & (7) & (8) & (9) & (10) & (11) & (12) & (13)\\
    \hline \input{catalogs/simple_table3.tex} \hline \hline
\end{tabular}
\caption{Structure of SWELLS {\it HST} Cycle 18 targets with
  multi-band imaging.  Col.~1 lists the lens ID;  Columns~2-9 the AB magnitudes of the bulge and the disc in the various bands, Col.~10 circularised half-light
  radius of the bulge; Col.~11 the axis ratio of the bulge; Col.~12
  circularised half-light radius of the disc; Col.~13 the axis ratio of
  the disc. The stellar masses inferred from these magnitudes are listed in Table~\ref{tab:sample4}.}
\label{tab:sample3}
\end{center}
\end{table*}

\begin{table*}
\begin{center}
  \scriptsize\begin{tabular}{ccccccccccccc} \hline \hline
    ID    & $\log_{10}(M_{\rm *,bulge}/\Msun)$ & $\log_{10}(M_{\rm

 *,disc}/\Msun)$  & $\log_{10}(M_{\rm *,bulge}/\Msun)$ & $\log_{10}(M_{\rm *,disc}/\Msun)$ \\
  &  Chabrier & Chabrier & Salpeter & Salpeter \\
   (1)   & (2)  & (3)  & (4)   & (5) \\
    \hline \input{catalogs/simple_table4.tex} \hline \hline
\end{tabular}
\caption{Stellar mass estimates for the SWELLS {\it HST} Cycle 18 targets with
  multi-band imaging.  Col.~1 lists the lens ID; Cols.~2 and 3 the
  stellar masses of the bulge and disc assuming a \citet{2003PASP..115..763C} IMF;
  Cols.~4 and 5 the stellar masses of the bulge and disc assuming a
  \citet{1955ApJ...121..161S} IMF.}
\label{tab:sample4}
\end{center}
\end{table*}

\begin{table*}
\begin{center}
  \scriptsize\begin{tabular}{ccccccccccccccccccc} \hline \hline
    ID    & $(M/L)_B$ & $(M/L)_V$ & $(M/L)_B$ & $(M/L)_V$ & $(M/L)_B$ & $(M/L)_V$ & $(M/L)_B$ & $(M/L)_V$  \\
  &  Bulge & Bulge & Disc & Disc & Bulge & Bulge & Disc & Disc \\
  &  Chabrier & Chabrier & Chabrier & Chabrier & Salpeter & Salpeter & Salpeter & Salpeter\\
   (1)   & (2)  & (3)  & (4)   & (5) & (6) & (7) & (8) & (9) \\
    \hline \input{catalogs/simple_table5.tex} \hline \hline
\end{tabular}
\caption{Stellar mass-to-light ratios, in solar units (i.e. $(M/M_\odot)/(L/L_\odot)$), for the SWELLS {\it HST} Cycle 18 targets with
  multi-band imaging.  Col.~1 lists the lens ID; Cols.~2-5 the
  mass-to-light ratio of the bulge and disc in the $B$ and $V$ bands assuming a \citet{2003PASP..115..763C} IMF;
  Cols.~6-9 the mass-to-light ratio of the bulge and disc in the $B$ and $V$ bands assuming a
  \citet{1955ApJ...121..161S} IMF.}
\label{tab:sample5}
\end{center}
\end{table*}

\section{Gravitational Lens Models}
\label{sec:grav}

In this section we describe the procedure used to obtain gravitational
lens models for each system.  We consider relatively simple mass
models for the deflector, described as a singular isothermal ellipsoid
plus external shear \citep{1994A&A...284..285K}. This simple model is known to provide a good description of
(at least early-type) galaxy-scale gravitational lenses \citep{2010ARA&A..48...87T, 2009ApJ...703L..51K, 2009MNRAS.399...21B}, and should provide sufficiently robust estimates of the physical quantities relevant for this present
analysis. That is, the mass enclosed within the critical curve, and the position angle and the
ellipticity of the mass distribution. The flat rotation curves of spiral galaxies \citep{1986RSPTA.320..447V} suggest that this model is appropriate for spiral lenses, at least to first order.

The modelling of the SWELLS lenses is complicated compared to typical
gravitational lens modelling, for several reasons. The first challenge
is the disentanglement of the light of the source and that of the
detector. Our strategy to address those is discussed in
\S~\ref{ssec:sub}. An additional challenge is the proper estimation of
model parameter uncertainties. This is discussed in \S~\ref{ssec:mod}
where we present our mass models and inference strategy. Finally, in
\S~\ref{ssec:ind} we discuss each individual system listing any
additional complexities.

\subsection{Disentangling the source and deflector light}
\label{ssec:sub}

The first step is the selection of the band which contains the highest
fidelity image of the lensed features, and will therefore be used to
produce the lens model. In most cases this is the bluest image
available, as the lensed features tend to be brightest at the blue end
of the spectrum.  In the cases where the lens galaxy contains
significant amounts of dust, a redder band was selected to minimise
complications from the dust.

Most significantly, the small Einstein radii of the lenses cause the
lensed features to lie in regions where the surface brightness of the
lens galaxy is non-negligible. This makes lens galaxy subtraction
non-trivial. In principle, we should fit the light profile of the lens
galaxy simultaneously with any lensing parameters and source surface
brightness profile parameters. We have attempted this, but found that
it was too computationally intensive to be practical.

In addition, the light profiles of the lens galaxies cannot be
described adequately by simple models such as the S\'ersic profile
\citep{Sersic1968} due to the presence of dust lanes, star forming
regions, and other complexities. Attempting to use such a profile
would result in lens galaxy subtractions with significant residuals
that would interfere with the modelling of the lensed features.

\subsubsection{Rotation Subtraction for Masking}

To proceed with the lens galaxy subtractions, we adopted the following
procedure. Firstly, for exploratory purposes, we take advantage of the
near-symmetry of the lens galaxy: by subtracting the image from a
180-degree rotated version of itself, the lensed features, often
asymmetric, are revealed. The centre point around which the rotation
takes place can be left as a free parameter, and optimised to find the
centre of rotation which makes the rotated image best match the
original image.

These rotation-subtracted images enable us to get an initial estimate
for the position and morphology of the lensed features, particularly
the identification of any faint counter-images. However, these images
cannot be used as data for modelling because they contain both
positive and negative reproductions of the arc.

\begin{figure}
\begin{center}
\includegraphics[scale=0.5]{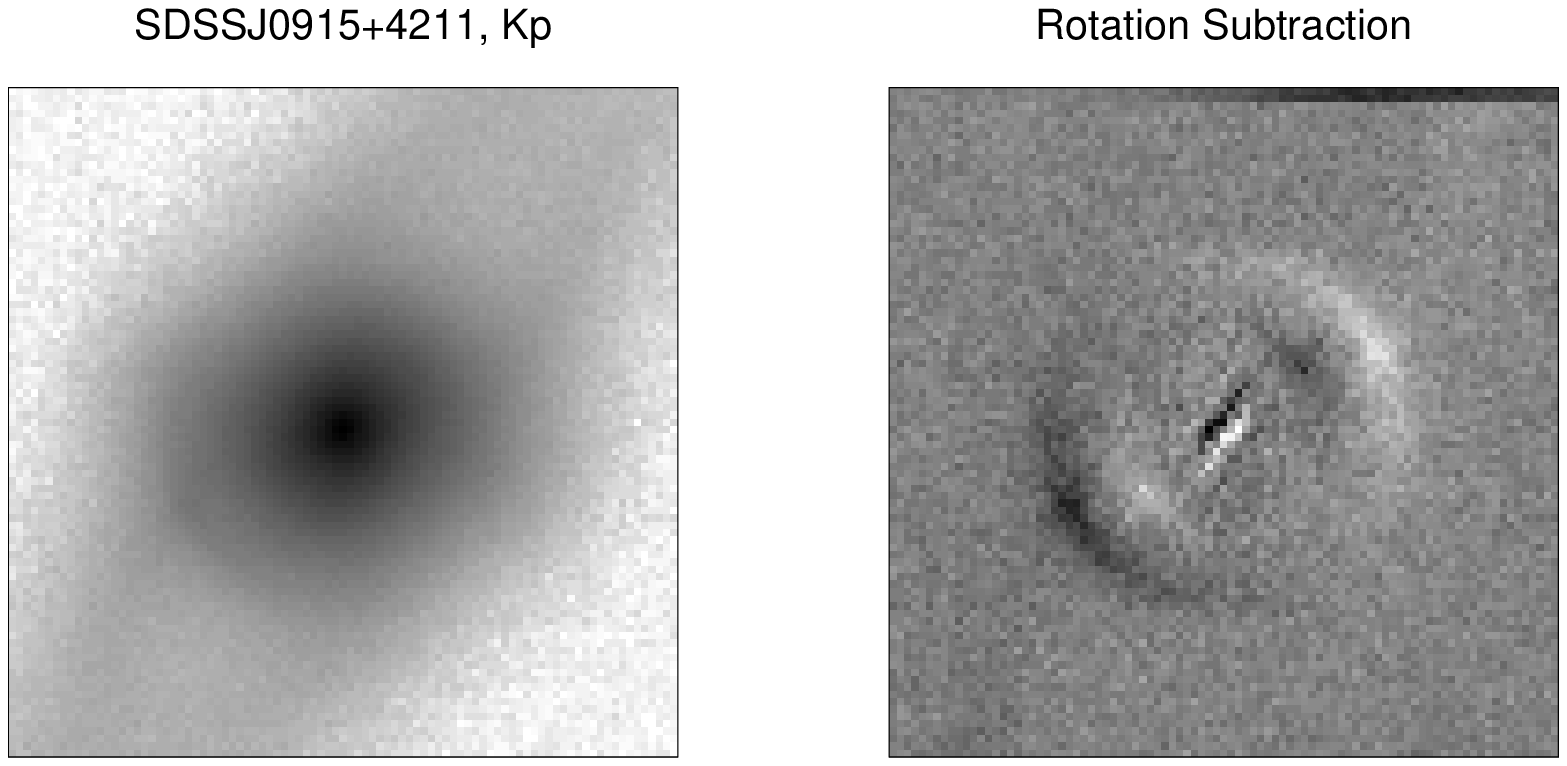}
\end{center}
\caption{Rotation subtraction for the Kp image of the SWELLS system
SDSSJ0915+4211. The lensed arc and counterimage are revealed by this
process. The rotation-subtracted image is then used to mask the
image so that a proper multi-Gaussian fit can be made.\label{0915_rotation}}
\end{figure}

In practice, we used the rotation-subtracted images to inform the
placement of a masked region, covering the lensed features. With
these features masked out, we fit a complex many-parameter model
for the surface brightness profile of the lens galaxy, and subtract
that model from the data, giving the final data for lens modelling.

\subsubsection{Multi-Gaussian Subtraction}

With the mask in place, we fit the remaining pixels using the code of
\citet{2011MNRAS.412.2521B}, with all S\'ersic indices set to $n=1/2$
--- i.e. we fit the lens galaxy with up to 10 elliptical Gaussian
components (this is similar to the Multi-Gaussian Expansion method of
\citet{Cappellari2002}). For the more complex galaxies with star
forming rings, we allow negative Gaussians to be included as well. The
star forming ring can be modelled as a smooth Gaussian-like profile
with negative Gaussians suppressing interior surface brightness,
resulting in a ring profile.  We use the multi-Gaussian model to
subtract the lens galaxy from the image, resulting in an image of the
lensed features only. These lens-subtracted images are shown in the
second column of Figure~\ref{lenses1}.

\subsection{Gravitational Lens Model}
\label{ssec:mod}

We fit Singular Isothermal Ellipsoid (SIE) lens models with external shear
to the lensed features of each system. The source plane
position $(x_s, y_s)$ and the image plane position $(x,y)$ of a ray
are related by: 
\begin{eqnarray}\label{lenseqn} x_s = x -
\alpha_x(x,y) \nonumber \\ y_s = y - \alpha_y(x,y) \end{eqnarray}
where the deflection angles $\alpha_{x,y}$ are given by
\citet{1998ApJ...495..157K}. Defining $\psi = \sqrt{q^2x^2 + y^2}$, where $q$ is the axis ratio/flattening of the projected mass profile,
the deflection angle formulae for the SIE component (i.e. before
adding the external shear) are:
\begin{eqnarray} \alpha_x(x,y) 
&=& \frac{b\sqrt{q}}{\sqrt{1-q^2}}\tan^{-1}\left[\frac{x\sqrt{1-q^2}}{\psi}\right]\\ \alpha_y(x,y) &=& 
\frac{b\sqrt{q}}{\sqrt{1-q^2}}\tanh^{-1}\left[\frac{y\sqrt{1-q^2}}{\psi + q^2 
}\right] \end{eqnarray}
as long as $q<1$. If $q>1$, then $q$ can simply be replaced by
$q^{-1}$ and the orientation angle $\theta$ rotated by 90 degrees,
with the same result. We do not allow $q=1$ but it can become
arbitrarily close to 1 and therefore close-to-spherical lenses are not
ruled out. The parameter $b$ denotes the Einstein Radius of the lens, using the intermediate-axis convention.

In many cases, we are attempting to fit this lens model to image
data that do not contain very much information. Therefore, informative
priors on the lens model parameters are required.
The centre of the lens model is fixed to be within $\pm$ 3 pixels of
the peak pixel of the light distribution. The position angle of the
lens model is assigned an informative prior, stating that the 
position angle of the lens is likely to be within $\pm 10$ degrees of
the position angle of the lens galaxy light. The modelling results in an
(intermediate axis) Einstein radius $b$ and an ellipticity parameter
$q$ for each system.  Once the lens model parameters have been
estimated, the Einstein Radius in radians can be converted to an
equivalent velocity dispersion $\sigma_{\rm SIE}$ using
\begin{equation}
\sigma_{\rm SIE} = c\sqrt{\frac{b}{4\pi}}\times\sqrt{\frac{D_{\rm s}}{D_{\rm ds}}}
\end{equation}
where $D_{\rm s}$ and $D_{\rm ds}$ are the angular diameter distances \citep{hogg}
from the observer to the source, and the deflector to the source, respectively.
For simplicity, we modelled the sources with a single circular
Gaussian component as this is adequate to explain the observed lensing
configuration in most cases. For those cases where this model was
inadequate, the source was generalised to be either an elliptical
Gaussian (e.g. for 0820) or multiple spherical Gaussians (e.g. for 0822). Due to the
presence of systematics in the data from incomplete subtraction of
lens galaxy light, and also the simplicity of the model used, we do
not literally perform the inference with the
\citet{2011MNRAS.412.2521B} likelihood $\mathcal{L}(\theta)$, as this
would result in inappropriately small uncertainties on the inferred
quantities due to the systematic errors. Simple, commonly adopted
models for the likelihood function tend to assume that the measurement
error in each pixel is independent, and that measurement error is the
sole cause of discrepancies between the model and the data. This is
not true in the presence of significant systematic effects such as
residuals from an incomplete lens galaxy subtraction, and the
assumption of a specific form (Gaussian) for the source light profile.

To account for this and
obtain realistic uncertainties on $b$ and $q$ despite the use of a simplified model, 
we adopted the following strategy. This strategy has
previously been briefly discussed by \citet{2011ApJ...733L..33B} and
\citet{2011MNRAS.412.2521B} and will be presented in depth in a forthcoming contribution (Brewer et al 2012, in preparation).

We carry out the sampling of the posterior distribution for the lens model parameters using a Nested Sampling algorithm \citep{skilling, 2009arXiv0912.2380B}. Rather than simply sampling the posterior distribution (which would yield unrealistic small uncertainties due to the presence of systematics and the simplified model), we explore the dependence of
the posterior distribution for the parameters $\theta$ on the {\it temperature}
$T$:
\begin{equation}
p(\theta | D; T) \propto \pi(\theta)\mathcal{L}(\theta)^{1/T}
\end{equation}
Where $\theta$ denotes the parameters of the lens model, $\pi(\theta)$
is the prior and $\mathcal{L}(\theta)$ is the standard likelihood
function. The temperature controls how conservative the posterior distribution is: $T = \infty$ reproduces the prior, and $T=1$ gives the standard posterior obtained by ignoring the systematics.

For each system, we then choose the highest temperature $T$
for which the posterior distribution over predicted lensed images (the 3rd panel in Figure~\ref{lenses1})
matches the gross lensed features seen in the data. In other words, we
force the posterior distribution to be as conservative as possible
while still fitting the
morphological features of the lensed images that are believed to be
robust. Note that using a temperature $T$ is
analogous to increasing the noise standard deviations on the data by a factor
$\sqrt{T}$. This procedure does not change the values of our best estimates for the lens
model parameters, but does increase the uncertainty around these estimates.
Not carrying out this procedure would result in substantially underestimated uncertainties.

See Figure~\ref{lenses1} for images
of the lenses, along with galaxy subtracted images, lens models
(chosen at random from the posterior distribution) and the residuals
of the lens model fits. In several cases there are significant
residuals remaining after the fitting procedure; these are generally
due to the source having more structure than the surface brightness
model that was assumed. In all cases the lensing aperture mass
estimates are reliable.

\begin{figure*}
\begin{center}
\includegraphics[scale=1.14]{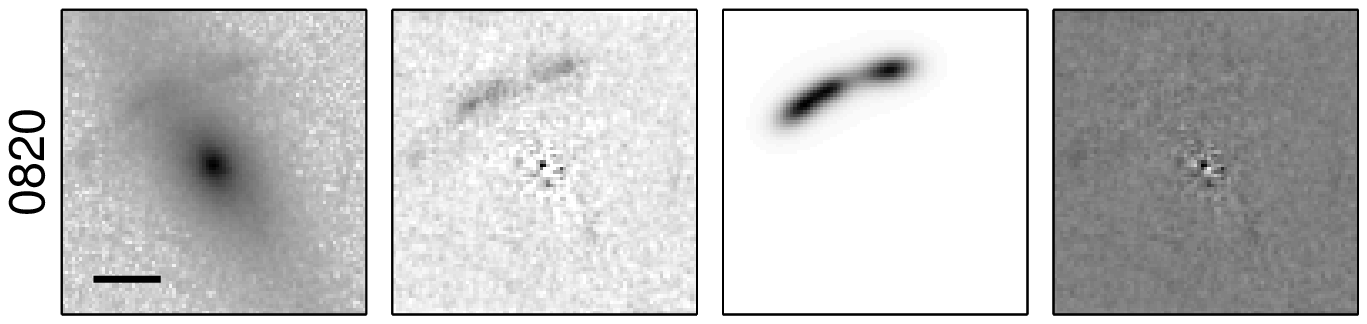}
\includegraphics[scale=1.14]{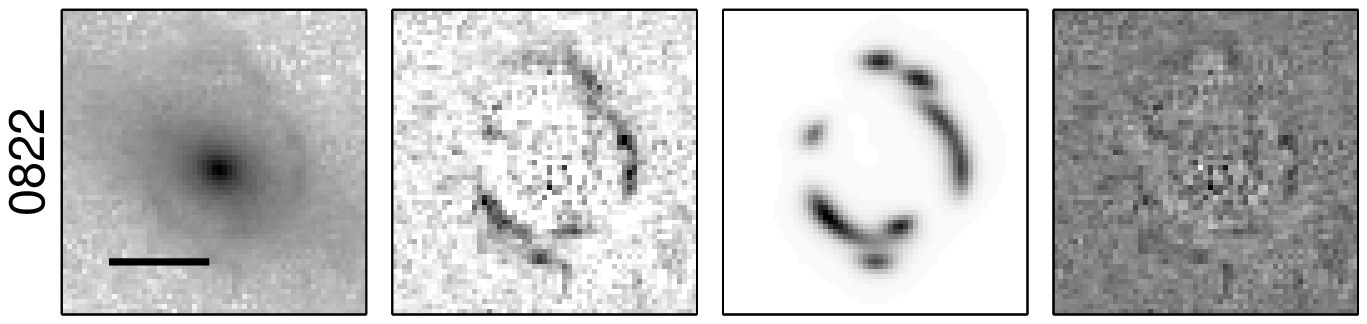}
\includegraphics[scale=1.14]{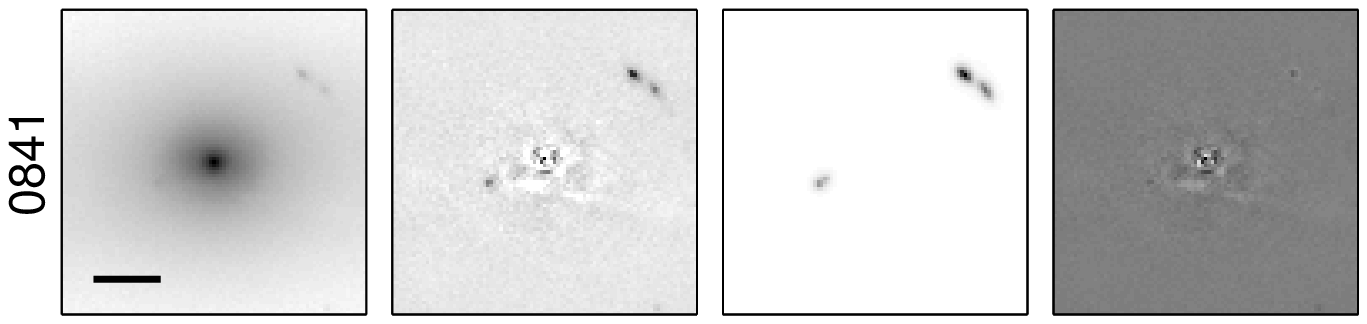}
\includegraphics[scale=1.14]{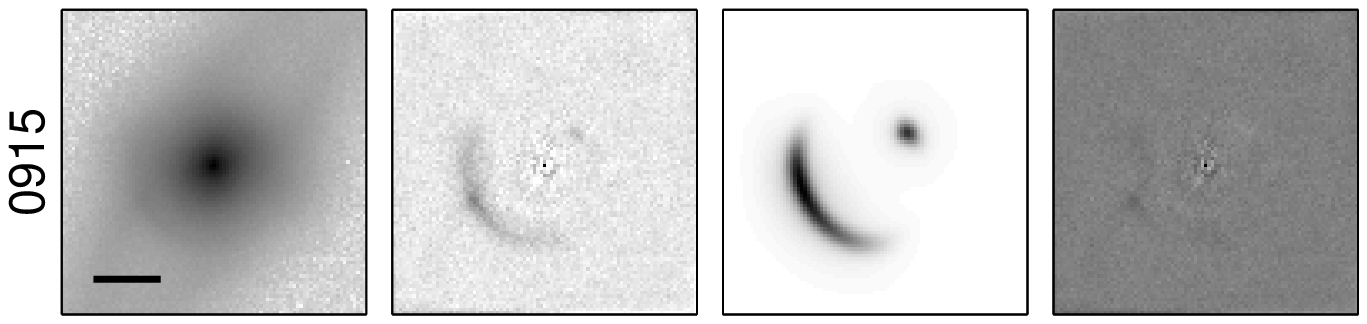}
\includegraphics[scale=1.14]{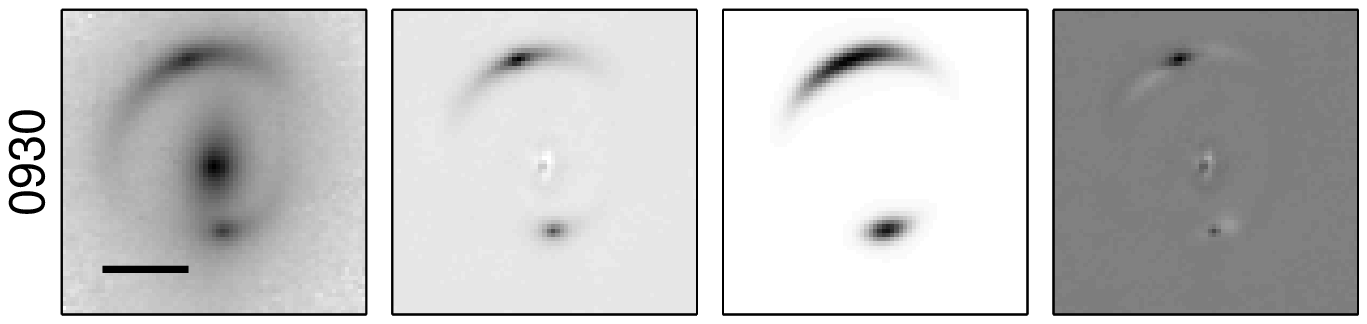}
\includegraphics[scale=1.14]{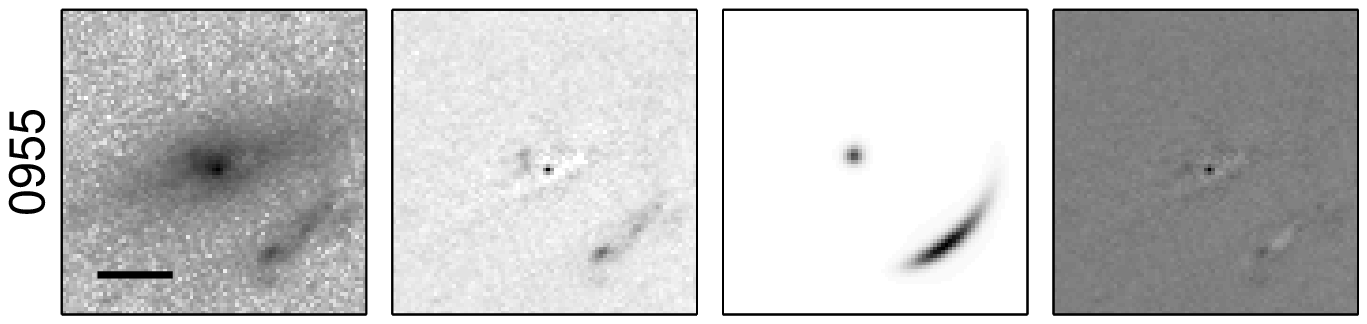}
\end{center}
\caption{Images and gravitational lens models of SWELLS systems. Left
  to right: Image, lens subtracted image, lens model predicted image (for a model chosen at random from the posterior distribution),
  residuals of lens model fit. The black bar in the left panel has a
  length of 1 arc second.\label{lenses1}} \addtocounter{figure}{-1}
\end{figure*}

\begin{figure*}
\begin{center}
\includegraphics[scale=1.14]{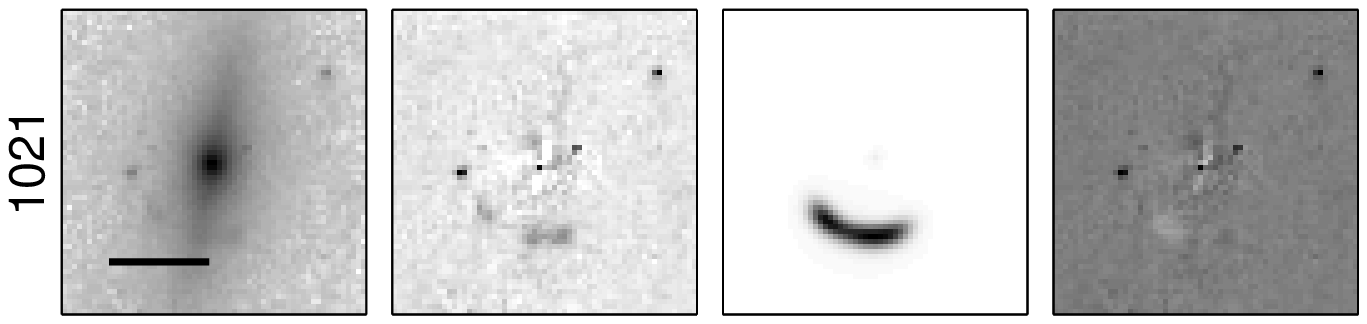}
\includegraphics[scale=1.14]{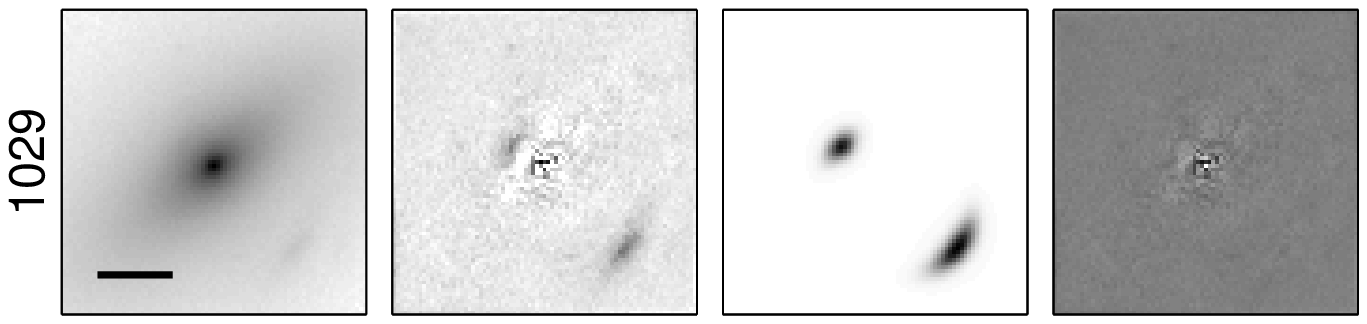}
\includegraphics[scale=1.14]{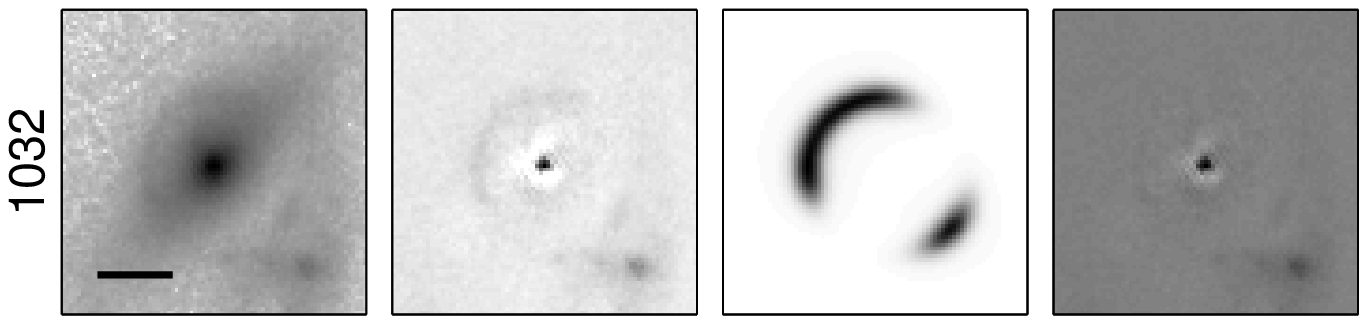}
\includegraphics[scale=1.14]{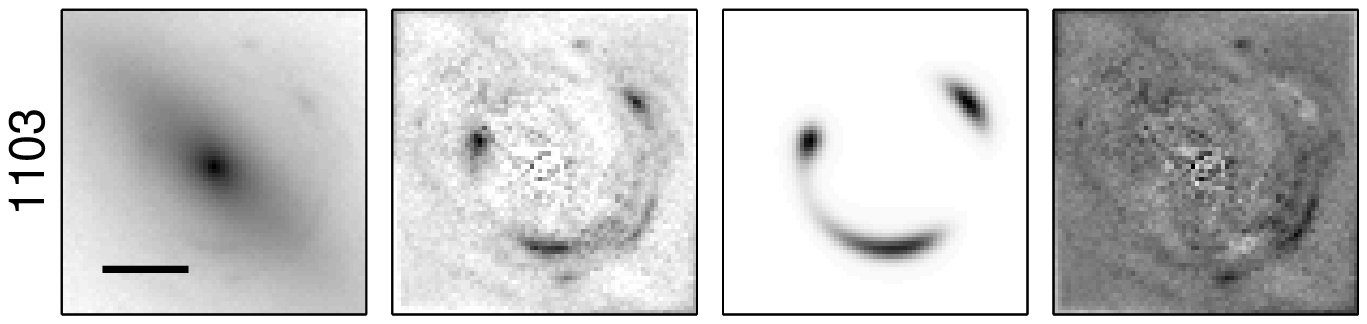}
\includegraphics[scale=1.14]{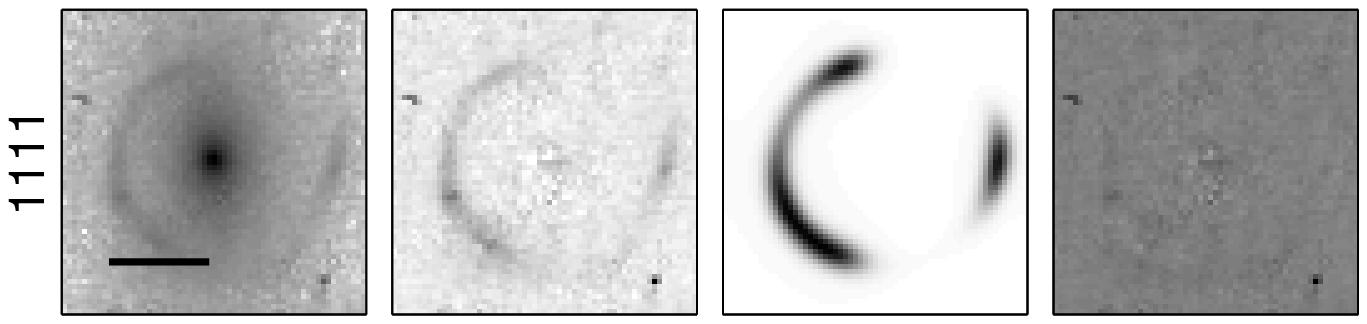}
\includegraphics[scale=1.14]{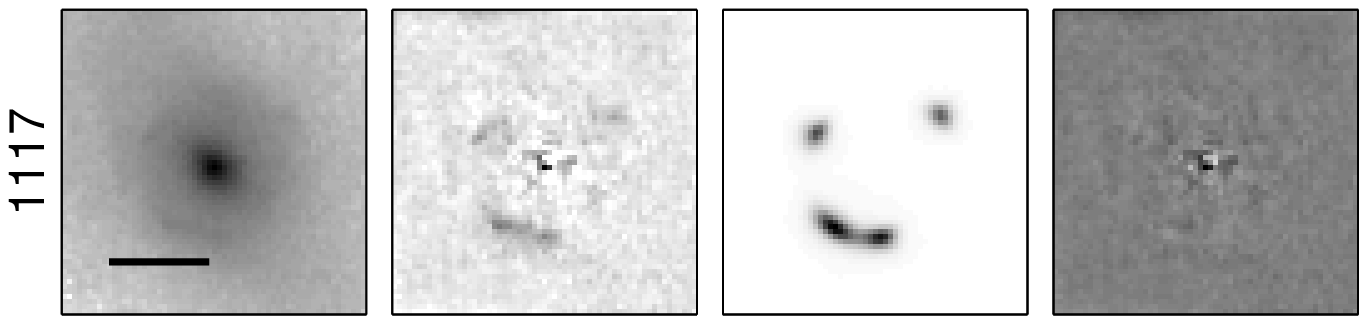}
\end{center}
\caption{(continued) Images and gravitational lens models of SWELLS systems. Left
  to right: Image, lens subtracted image, lens model predicted image (for a model chosen at random from the posterior distribution),
  residuals of lens model fit. The black bar in the left panel has a
  length of 1 arc second.\label{lenses2}} \addtocounter{figure}{-1}
\end{figure*}

\begin{figure*}
\begin{center}
\includegraphics[scale=1.14]{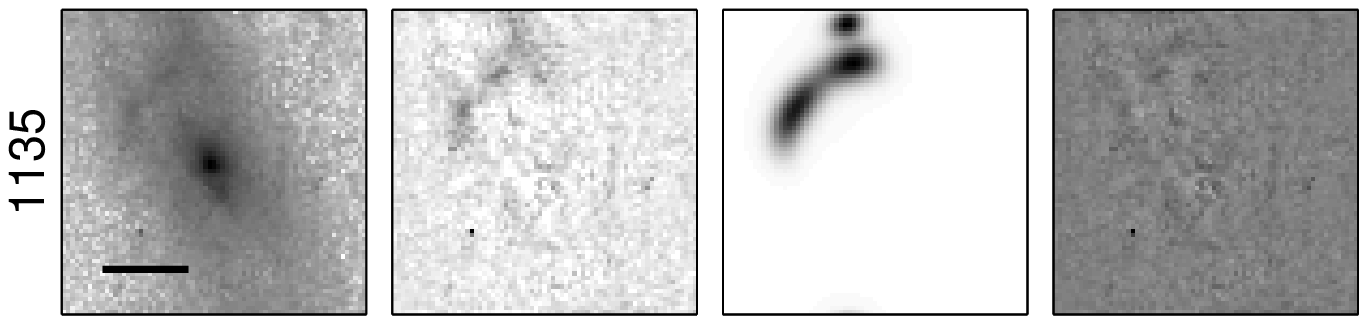}
\includegraphics[scale=1.14]{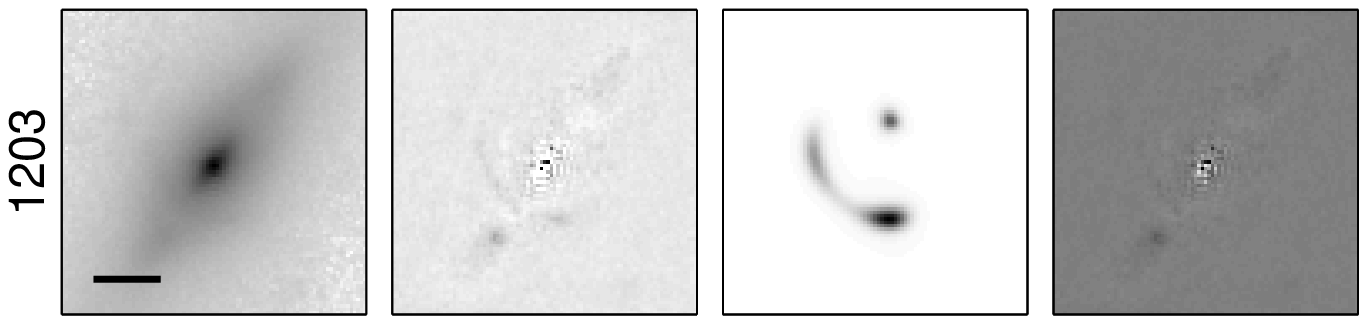}
\includegraphics[scale=1.14]{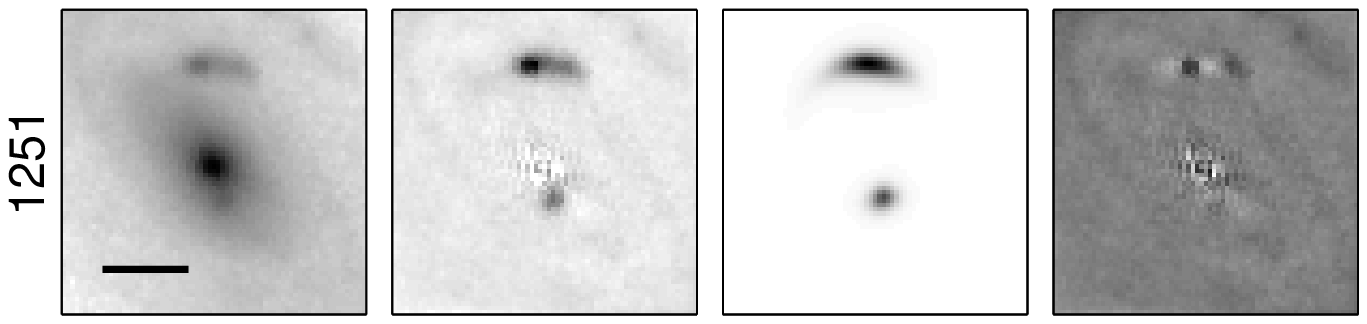}
\includegraphics[scale=1.14]{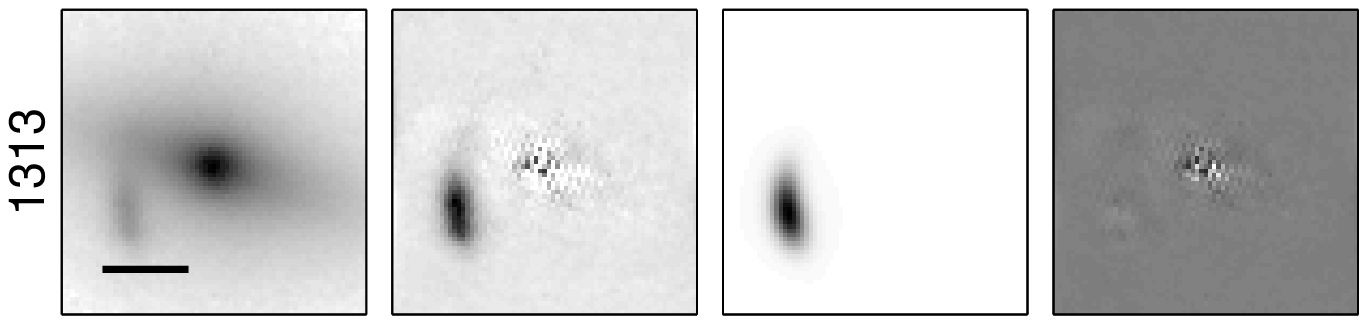}
\includegraphics[scale=1.14]{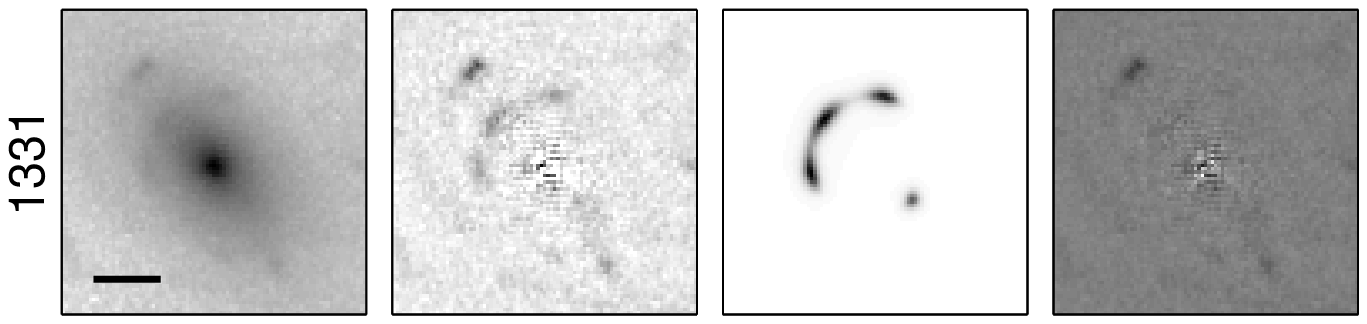}
\includegraphics[scale=1.14]{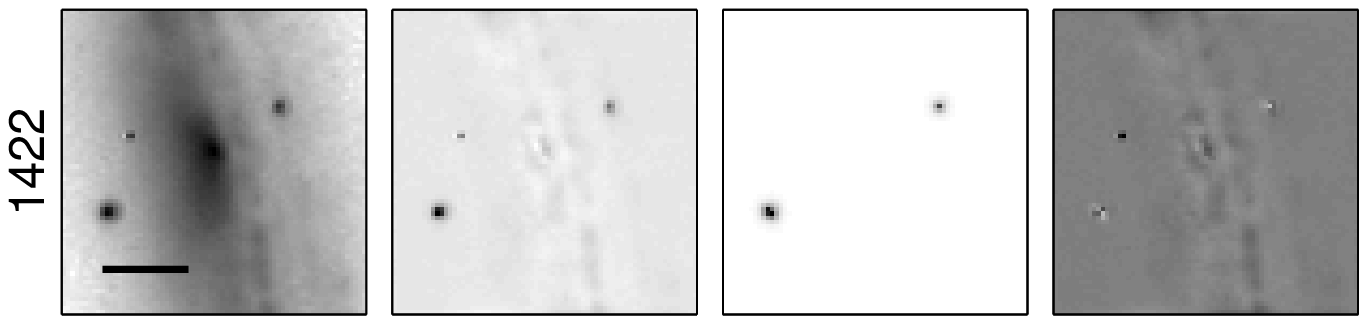}
\end{center}
\caption{(continued) Images and gravitational lens models of SWELLS systems. Left
  to right: Image, lens subtracted image, lens model predicted image (for a model chosen at random from the posterior distribution),
  residuals of lens model fit. The black bar in the left panel has a
  length of 1 arc second.\label{lenses3}} \addtocounter{figure}{-1}
\end{figure*}

\begin{figure*}
\begin{center}
\includegraphics[scale=1.14]{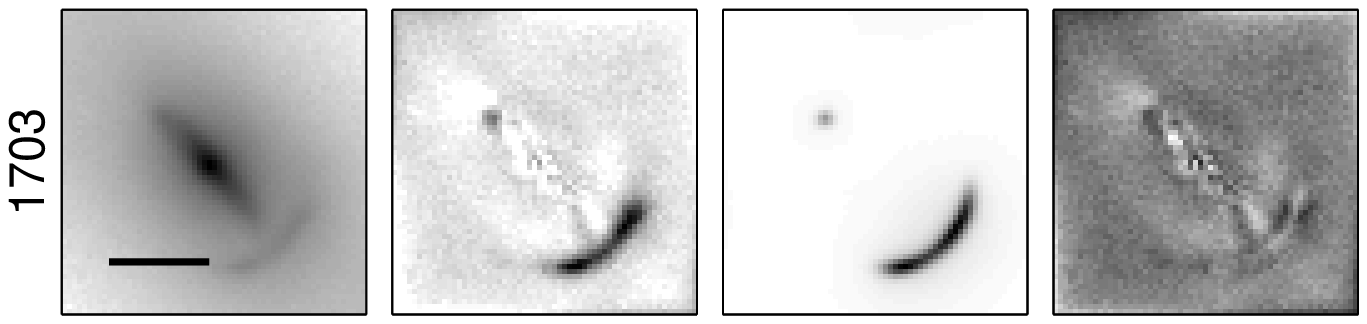}
\includegraphics[scale=1.14]{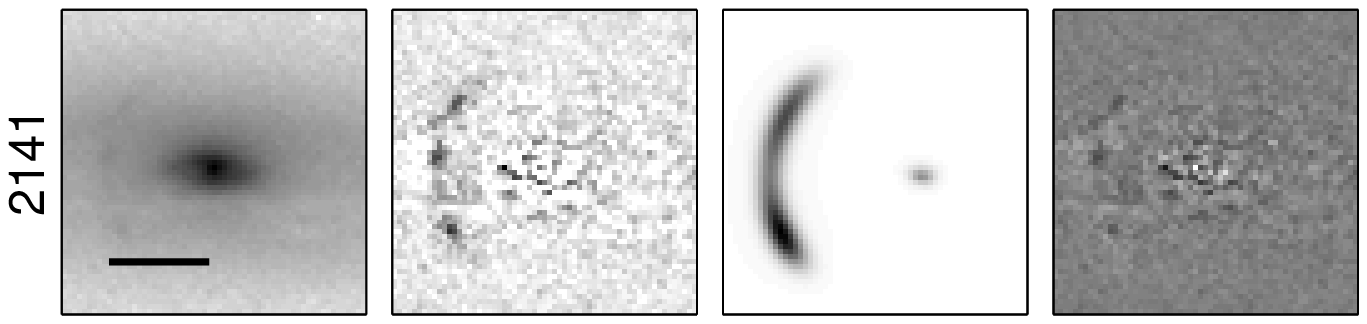}
\end{center}
\caption{(continued) Images and gravitational lens models of SWELLS systems. Left
  to right: Image, lens subtracted image, lens model predicted image (for a model chosen at random from the posterior distribution),
  residuals of lens model fit. The black bar in the left panel has a
  length of 1 arc second.\label{lenses4}}
\end{figure*}

\subsection{Notes on Individual Systems}
\label{ssec:ind}

In this section we give brief notes on the image configurations of the
lens models.

{\bf 0820+4847} Very little curvature is present in the arc. No
definite counter-image is detected. An elliptical source is required,
and the deflector is inferred to be very flat. 

{\bf 0822+1828} Multiple (three) source components are required. This
results in tight constraints on the lens model parameters.

{\bf 0841+3824} Two source components are required. Whereas a bimodal
posterior PDF is found, one mode chosen because of unrealistic small
axis ratios for the other modes. A consistent lens model for this system has also
been presented by B08.

{\bf 0915+4211} Straightforward fit of the arc and counterimage with a
single circular source.

{\bf 0930+2855} Significant residuals due to the assumptions of no shear and
isothermality (Auger et al., in preparation). The source has very high
surface brightness and the system is an example of an early-type/early-type
lensing system \citep[e.g.,][]{auger2011}.

{\bf 0955+0101} Straightforward fit of arc and counterimage with a
double image configuration. A consistent lens model for this system has also
been presented by B08.

{\bf 1021+2028} We fit to the arc only. A bimodal solution was found,
but we could rule out the mode which did predict a distant unseen
counter-image.

{\bf 1029+0420} Classic double image configuration. A consistent lens model for this system has also
been presented by B08.

{\bf 1032+5322} Arc and counterimage are well modelled by a single
circular source. A consistent lens model for this system has also
been presented by B08.

{\bf 1103+5322} Arc can be explained by a single circular source
imaged four times. Lens model parameters are tightly constrained.
A consistent lens model for this system has also
been presented by B08.

{\bf 1111+2234} Arc can be explained by a single circular source
imaged four times. Lens model parameters are tightly constrained.

{\bf 1117+4704} The single circular source is lensed into a quad
configuration. Lens model parameters are tightly constrained.

{\bf 1135+3720} The arc structure is complex, requiring two components
for a good fit. Position of the counterimage is uncertain. Lens
parameter constraints are not particularly tight.

{\bf 1203+2535} Position and morphology of the arc can be reproduced
by a single circular source. Subtraction of the deflector galaxy was
not straightforward for this system, leaving residuals that may add
systematic uncertainties.

{\bf 1251-0208} Classic double image configuration. A consistent lens
model for this system has also been presented by B08.

{\bf 1313+0506} Arc is well matched by a lensed circular source, but
due to the non-detection of a counter-image, the constraints on the
lens model parameters are weak.

{\bf 1331+3638} Classic cusp configuration leads to strong constraints
on the lens model parameters.

{\bf 1422+4134} This is the only grade 'B' lens system for which a
plausible model has currently been obtained. The two compact images
are possibly double images of a background QSO. This configuration
yields weak constraints on the lens model, and the inferred stellar
mass fractions place this system as a significant outlier from the
relation in Figure~\ref{correlation1}, possibly due to its significant
dust content or a potential misidentification of the source
redshift. Hence, this system has been excluded from the IMF analysis
of Section~\ref{imf}.

{\bf 1703+2451} Arc and counterimage can be explained by a single
circular source.

{\bf 2141-0001} As discussed in paper II, the arc appears to be split
into three images, but this is likely due to the presence of two dust
lanes. The arc morphology can be explained by a single circular
source.  The posterior distribution includes models where the arc is
split into distinct images, as well as models where the arc is
continuous.


\section{Comparing Stellar Mass and Lensing Mass}
\label{imf}

In this section we derive lensing and stellar masses within the
critical curves (\S~\ref{ssec:masses}), and compare them to derive
limits on the stellar in initial mass function (\S~\ref{probabilistic}).

\subsection{Lensing and stellar masses within the critical curves}
\label{ssec:masses}

As in Paper I, we fit the surface brightness
models to the available high resolution imaging, i.e. the HST and Keck-AO imaging, using the multi-colour fitting code \SPASMOID \citep{2011ApJ...726...59B}.
The stellar mass (and surface brightness)
model consists of an $n=4$ S\'ersic profile (i.e. a de Vaucouleurs
profile) for the bulge and an $n=1$ exponential disc component. One advantage of \SPASMOID over
similar methods such as {\sc GALFIT} \citep{2002AJ....124..266P} include the ability to take into account prior information such as the fact that the disc component should have a larger effective radius and a higher ellipticity than the bulge. Once we have obtained the surface photometry parameters, we then infer the stellar masses of the bulge and disc components using a Bayesian method for fitting stellar population synthesis models to multi-band photometric data \citep{2009ApJ...705.1099A}. This method uses a grid of models output from a standard SPS code, and allows for marginalisation over unknown parameters such as star formation history, metallicity, etc. A
summary of the structural properties (stellar masses, sizes, axis
ratios) of the SWELLS Cycle 18 targets with multi-band imaging is
given in Tables~\ref{tab:sample3} and~\ref{tab:sample4}. For the other lenses we use the
structural parameters as reported in SWELLS I
\citep{2011MNRAS.417.1601T}.

We then integrate these models to obtain the total stellar mass
$M_{*,\rm SPS}$ within the critical curve. This can be compared with
the inferred mass within the critical curve from lensing ($M_{\rm
  lens}$), yielding stellar mass fractions $f^* = M_{*,\rm SPS}/M_{\rm
  lens}$ under both the Chabrier IMF assumption and the Salpeter IMF
assumption. The reason for the choice of the critical curve as the
integration aperture is that, heuristically, it is the aperture within
which the mass of the lens is most strongly constrained
\citep[e.g.,][and references therein]{2010ARA&A..48...87T}. To obtain
uncertainties on the stellar mass fractions, we Monte-Carlo sample
from the posterior distribution for the lens model parameters and the
posterior distributions for the total stellar mass.

\begin{figure*}
\vspace{0.5cm}
\includegraphics[scale=0.7]{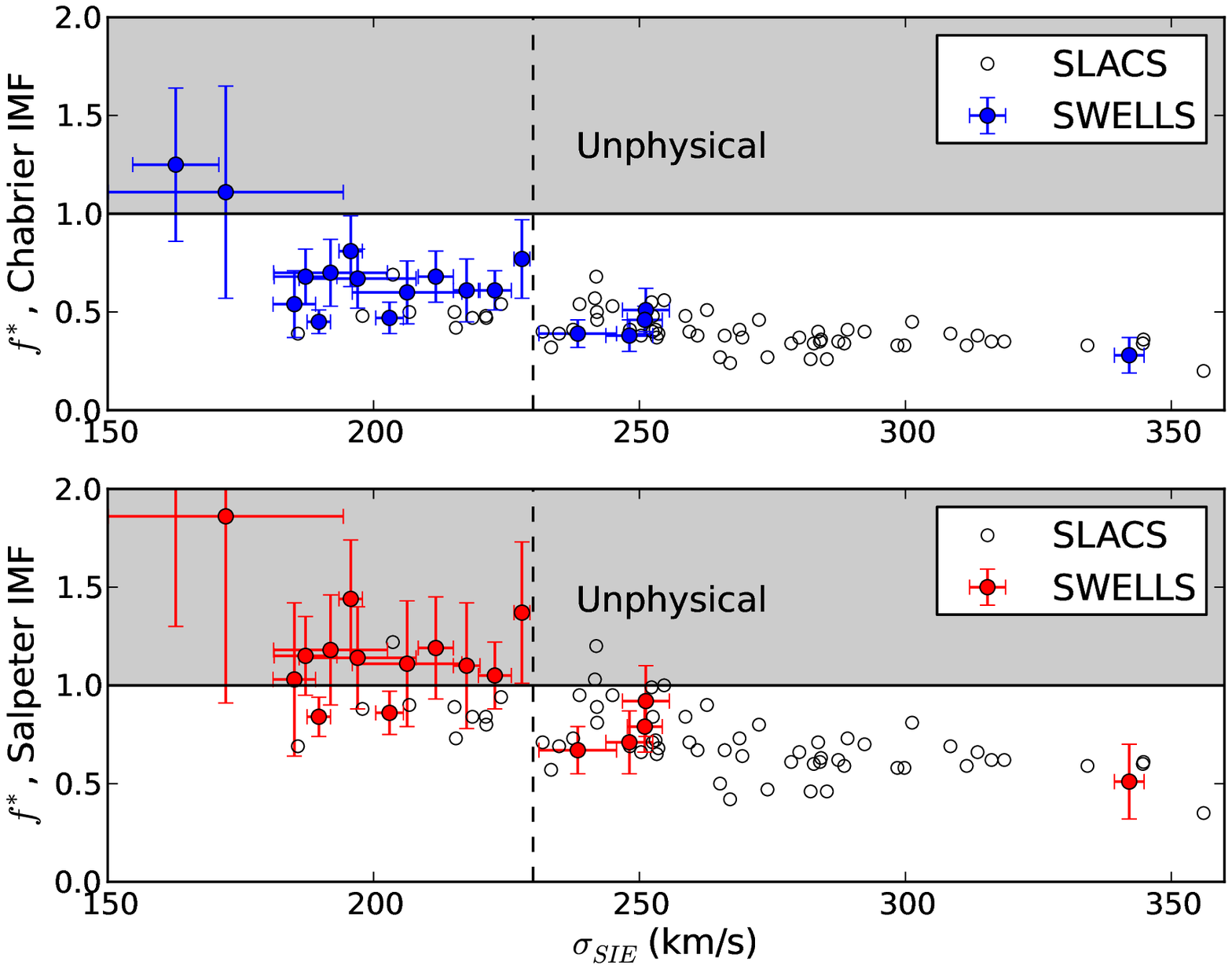}
\caption{Stellar mass fraction within the critical curve vs lensing
  velocity dispersion for the SWELLS sample. The SLACS sample (with
  duplicates in SWELLS removed) \citep{2009ApJ...705.1099A} is also
  plotted for comparison. Note that, at the low mass end (to the left of the dashed line at $\sigma_{\rm SIE} < 230 \kms$), the Salpeter IMF assumption results in inferred stellar mass fractions greater
  than unity, which is unphysical.
\label{correlation1}}
\end{figure*}

The resulting stellar mass fractions for the grade 'A' lenses, under
the Chabrier and Salpeter assumptions, are plotted in
Figure~\ref{correlation1} as a function of lensing velocity dispersion
(a proxy for overall galaxy mass that can be measured accurately and
independent of the size of the Einstein Radius and the normalisation
of the stellar IMF).  Remarkably, there is a smooth declining trend as
a function of stellar velocity dispersion over more than a factor of
two in stellar velocity dispersion (and a factor of $\sim$10 in
stellar mass). This trend can be due to two effects: i) a systematic
variation of the IMF; ii) a varying dark matter fraction within the
Einstein Radius due to a change in stellar mass to dark matter ratio and the
change in physical size of the Einstein radius. Disentangling the two
effects is a major goal of the SWELLS survey and requires very
realistic lensing and dynamical models (e.g. paper II). However, as we
will show in the rest of this paper, a very robust inference on the
normalisation of the IMF at the low velocity dispersion end can be
obtained based on a very simple argument.

At the low velocity dispersion end, the Salpeter IMF assumption
results in inferred stellar mass fractions greater than unity, which
is clearly unphysical, especially considering that there may be an
additional contribution of cold gas distributed as the stars (paper II
and references therein). This provides evidence against the Salpeter
IMF and any IMF that would predict a similar or greater stellar mass.  In Section~\ref{probabilistic} we formalise this
argument with a probabilistic model and derive quantitative
constraints on the IMF for the lower mass ($\sigma_{\rm SIE} < 230$ km
s$^{-1}$) SWELLS galaxies.

\subsection{Constraints on the Initial Mass Function}
\label{probabilistic}

Our data do not allow us to probe the form of the IMF but they do
allow us to place robust constraints on the integral of the IMF: if
the assumed IMF yields a mass significantly greater than the inferred
total mass from lensing (i.e., a stellar mass fraction much greater
than unity) then this IMF is excluded by our data. We could assume
that all of our galaxies have the same underlying central stellar mass
fraction $f^*_{\rm true}$, in which case the estimate for a Salpeter
IMF (i.e., the variance-weighted mean of the points in
Figure~\ref{correlation1}) would imply $f^*_{\rm true} > 1$. However,
there is certainly some amount of scatter in the central dark matter
fraction and therefore in the stellar mass fraction.  We therefore
explicitly build a model for the true underlying stellar mass fraction
that includes Gaussian scatter and the fact that $f^*_{\rm true}$ is
between 0 and 1. In practice this is a truncated normal distribution,
which we parameterise as having a peak at $\mu$ and characteristic
width $\sigma$. Then the probability of observing a set of galaxies
with true stellar mass fractions \{$F^*$\} is given by
\begin{equation}
\label{equation_Ftrue}
p(\{F^*_i\} | \mu, \sigma) = \prod_{i=1}^n \frac{\exp\left[-\frac{1}{2\sigma^2}(F^*_i - \mu)^2\right]}{\int_0^1\exp\left[-\frac{1}{2\sigma^2}(F^*_i - \mu)^2\right] \, dF^*_i}.
\end{equation}
where $F^*_i \equiv f^*_{\rm true, i} \in [0, 1]$. We also introduce a parameter
$\alpha$ that parameterises the difference between a true (unknown)
normalisation of the IMF and the normalisation resulting from a
Salpeter IMF:
$$
M_{*,\rm{true}} = \alpha M_{*,\rm{Salp}}.
$$
This normalisation is always greater than 0, and $\alpha = 1$ implies a
Salpeter IMF (a Chabrier IMF typically would imply $\alpha \sim 0.55$).
At fixed total mass, an increase in $\alpha$ yields an increase in
$f^*_{\rm true}$, demonstrating that the parameters are highly
degenerate. However, as noted previously, $\alpha$ cannot be raised
to arbitrarily large values because the stellar mass fraction is bounded
to be less than unity. In practice, $\alpha$ plays a similar role to a
stellar mass-to-light ratio but has the benefit of being less dependent on dust
and age due to having explicitly modelled the spectral energy distributions
of the galaxies (see Section~\ref{ssec:masses}).

Given our data -- total masses from lensing and stellar masses
assuming a Salpeter-like normalisation of the IMF -- and invoking
Bayes' theorem allows us to write down the probability distribution
for our unknown model parameters $\mu$, $\sigma$, $\alpha$ and
$\{F^*_i\}$, given the data:
\begin{eqnarray}
p(\alpha, \mu, \sigma, \{F^*_i\} | \{f^*_i\}) &\propto& p(\alpha, \mu, \sigma, \{F^*_i\}) \nonumber \\
& &  \times p(\{f^*_i\} | \alpha, \mu, \sigma, \{F^*_i\}) \\
&=& p(\alpha)p(\mu)p(\sigma) \nonumber \\
& & \times p(\{F^*_i\} | \mu, \sigma)p(\{f^*_i\} | \{F^*_i\})
\end{eqnarray}
where $f^*_i = \alpha f^*_{i,{\rm Salp}}$ are the scaled observed
stellar mass fraction, $\{F^*_i\}$ is the set of true (unknown)
stellar mass fractions, and the leading terms are priors on $\alpha$,
$\mu$, and $\sigma$.  The term $p(\{F^*_i\} | \mu, \sigma)$ is given
by Equation \ref{equation_Ftrue} and the term $p(\{f^*\} | \{F^*_i\})$
is the likelihood of having observed the data, taken to be a
multivariate normal distribution with means $\{F^*_i\}$ and standard
deviations given by the measurement uncertainties (the galaxies are
independent and there are therefore no covariances).

The inferences on the parameters ($\mu$, $\sigma$, $\alpha$) were
computed using Markov Chain Monte Carlo, with the true $F^*$s
explicitly integrated out:

\begin{eqnarray}
p(\alpha, \mu, \sigma, | \{f^*_i\}) &=& \int p(\alpha, \mu, \sigma, \{F^*_i\} | \{f^*_i\}) \, d^n F^*_i
\end{eqnarray}

\begin{figure}
\vspace{0.5cm}
\includegraphics[scale=0.47]{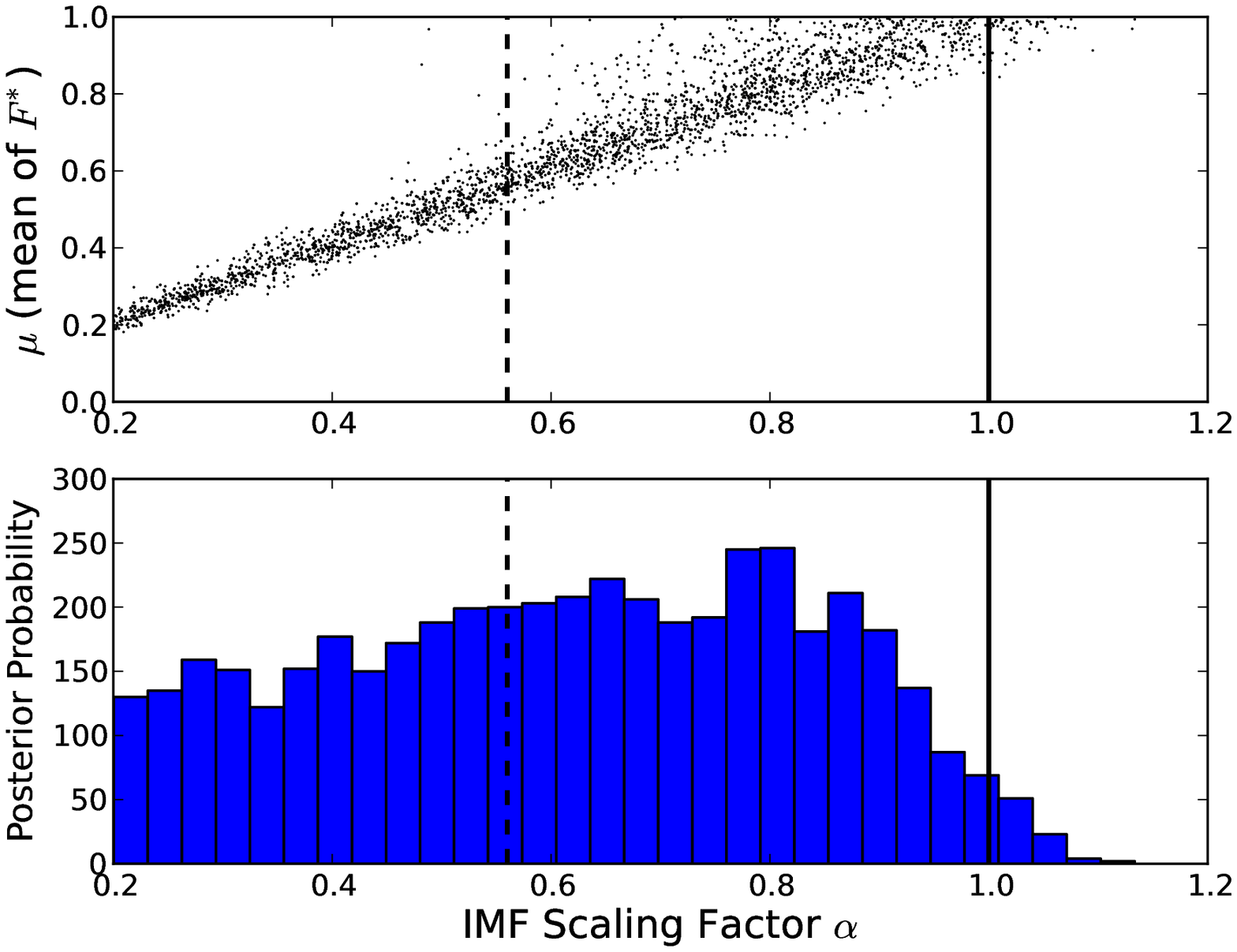}
\caption{Inference on the IMF mismatch parameter $\alpha$ and the
  parameter $\mu$, the mean of the true distribution of stellar mass
  fractions. Values of $\alpha$ less than about 0.9 are approximately
  equally plausible, but the evidence strongly disfavours higher
  values of $\alpha$. The vertical line denotes the Salpeter IMF, and
  the dashed line indicates a Chabrier-like (in terms of mass-to-light
  ratio) IMF. The probability that the IMF is lighter than Salpeter is
  98\%.\label{inference1}}
\end{figure}

\begin{figure}
\vspace{0.5cm}
\includegraphics[scale=0.47]{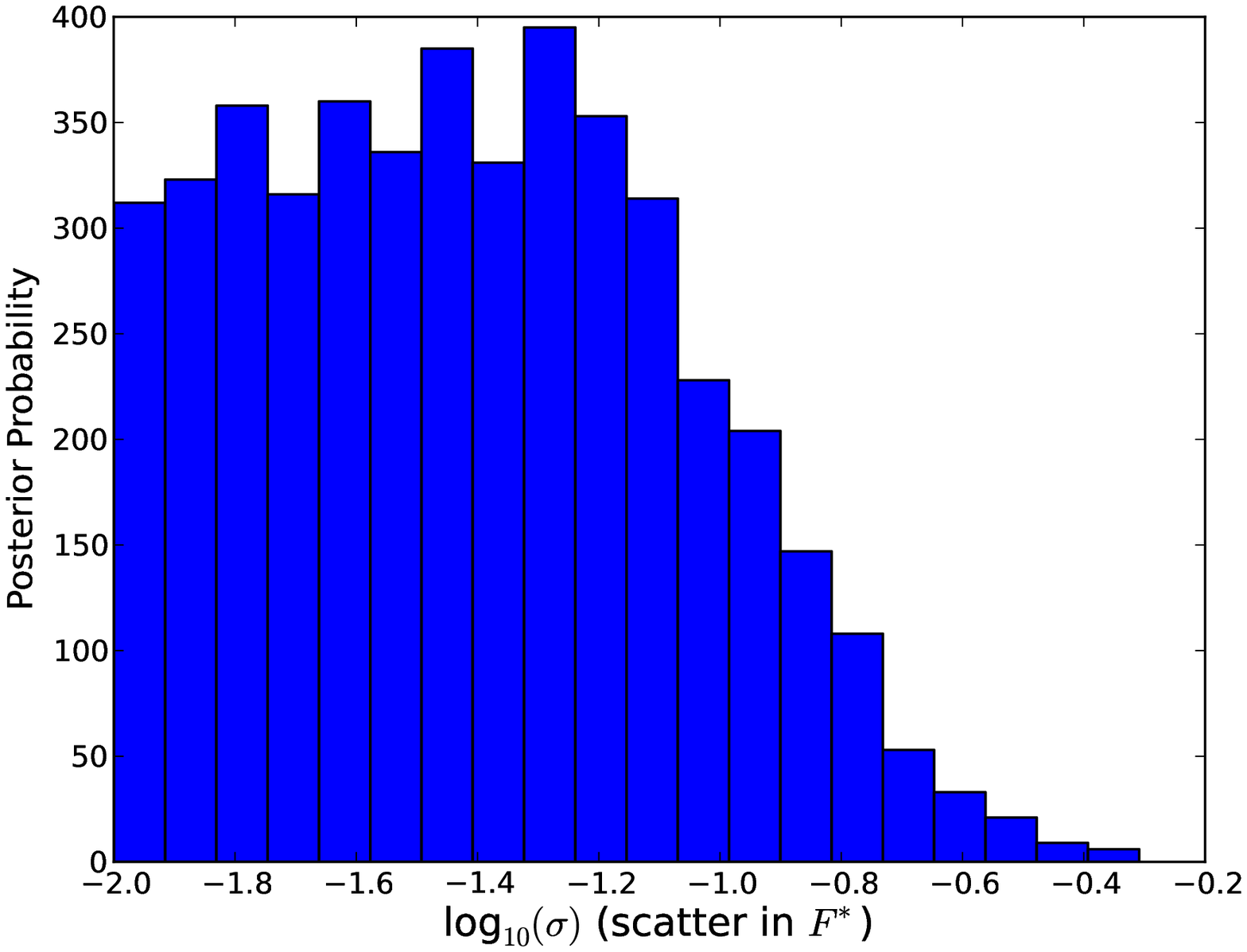}
\caption{Inference on the parameter $\sigma$, the intrinsic scatter in
  the true distribution of stellar mass fractions $F^*$. The value of
  $\sigma$ is constrained to be low ($< 0.16$ with 95\% probability),
  so there is evidence that the intrinsic stellar mass fractions are
  fairly homogeneous across the low mass end of the SWELLS
  sample.\label{inference2}}
\end{figure}

These inferences were calculated assuming a uniform prior on $\mu$ (a
location parameter) between 0 and 1, a log-uniform prior on $\sigma$
(a scale parameter) between 0.01 and 1, and a log-uniform prior on
$\alpha$ (a scale parameter) between 0.2 and 5, all independent.  The
results are shown in Figures~\ref{inference1}
and~\ref{inference2}. Values of $\alpha$ less than about 0.9 are
approximately equally plausible, but the evidence strongly disfavours
values of $\alpha$ greater than unity. The evidence for the point
hypothesis $\alpha=0.5$ is only about twice that of the point
hypothesis $\alpha=1$, however the probability that the IMF is lighter
than Salpeter is 98\%. This result depends slightly on the chosen
prior for $\mu$, $\alpha$ and $\sigma$, and becomes weaker if a
greater prior probability is implicitly assigned to the hypothesis
that the intrinsic $F^*$ values are clustered close to 1. For example,
if the prior bounds on $\mu$ are extended to $-0.1$ and $1.1$,
allowing the true $F^*$ distribution to place most of its mass near 1,
the posterior probability that the IMF is lighter than Salpeter is
slightly reduced to 93\%.  The intrinsic scatter on the true $F^*$
values is inferred to be small ($\sigma < 0.16$ with 95\% probability).

Note that, if we take into account the fact that there can be an
additional 20\% of mass in cold gas distributed in the same way as the
stars (as in Papers II), the probability that the IMF is lighter than
Salpeter becomes greater than 99\%. Correspondingly, the 95\%
probability upper limit to $\alpha$ becomes 0.80.

\section{Discussion and comparison with previous work}
\label{sec:disc}

The main result of this paper is that the stellar mass of spiral
galaxies with lens velocity dispersion between 160 and 230 $\kms$
(i.e. with masses comparable to or slightly above the Milky Way)
is lighter than that predicted based on their colours using a Salpeter
IMF.

This conclusion is based on a robust and straightforward argument,
that is that the stellar mass cannot exceed the total mass measured by
strong gravitational lensing. A high abundance of cold gas would make
our result even stronger, requiring even less mass in stars and
therefore a lighter normalisation of the IMF. Systematic uncertainties
in the stellar mass estimates due to differences in stellar population
synthesis models and assumptions are of the order of 0.05-0.1 dex for
a fixed IMF \citep{2009ApJ...699..486C,
  2009ApJ...705.1099A,2010ApJ...709.1195T} and therefore not
sufficient to modify our conclusions significantly.

Even though our sample is the largest sample of spiral lens galaxies
currently available it is important to consider whether this result
could somehow be biased by unknown selection effects. In paper I we
investigated the selection function in detail by comparing the size
mass relation of SWELLS lenses to that of non-lens galaxies selected
from SDSS and found them to be statistically equivalent. The detailed
analysis of the lens SDSSJ2141-0001 presented in paper II indicate
that that galaxy might be drawn from the densest part of the
distribution of mass density profiles of spiral galaxies. Whether this
is a generic feature of lensing selection needs to be verified by the
detailed analysis of the entire sample. However, if lens galaxies were
representative of the densest spiral galaxies, this would only make
our inference stronger with respect to the overall population, since
the densest galaxies are likely those with the highest stellar
content.

Our results are consistent with independent estimates of the
normalisation of the stellar IMF of spiral galaxies based on a variety
of techniques. For example, based on spiral galaxy rotation curves,
\citet{2001ApJ...550..212B} find that a Salpeter IMF is too heavy and
argue for a normalisation that is significantly lower. Based on the
vertical velocity dispersion of discs, the Disk Mass survey
\citep{2011ApJ...739L..47B,2010ApJ...716..198B} argues that the
mass-to-light ratio (and hence hence the normalisation of the stellar
IMF) is significantly lower than that inferred assuming a Salpeter IMF.
Interestingly, a qualitatively similar result is found for elliptical
galaxies in the same velocity dispersion range. Dynamical arguments
\citep{2006MNRAS.366.1126C,2011arXiv1111.2905D} as well as joint
dynamical and lensing arguments
\citep{2010MNRAS.406.2339B,2011MNRAS.415.2215B} show that a Salpeter
IMF is too heavy for low mass ellipticals.

The light normalisation of the stellar IMF (or equivalently
mass-to-light ratio) required for spirals and ellipticals comparable
to the mass of the Milky Way is in contrast with the heavier
normalisation required for more massive systems. For example,
\citet{2010Natur.468..940V}, \citet{2011MNRAS.417.3000S},
\citet{2010ApJ...721L.163A} and \citet{2011arXiv1111.4215S}, using
different techniques and datasets, find that light Chabrier-like IMFs
are ruled out for the most massive early-type galaxies, and a heavier
IMF is preferred. Further evidence for this hypothesis is provided by
\citep{2010ApJ...709.1195T,2011MNRAS.415..545T} -- and
\citet{2011ApJ...728L..39N} for the case of massive brightest cluster
galaxies -- although those studies could not break completely the
degeneracy between dark matter content and normalisation of the IMF.

Taken together, these results are consistent with the hypothesis that
the stellar IMF is not universal, but that it changes systematically
with galaxy stellar velocity dispersion and mass (and therefore halo
mass to first approximation). Since stellar velocity dispersion
correlates with the age of the stellar populations, this
non-universality might reflect the non-universality of the conditions
that were prevalent when the stars were formed. The obvious suspects
are the density of the gas, and the condition of ambient background
radiation. A key open question is whether this variation in the IMF
can be understood theoretically based on first principles.

It is important to emphasise that lensing and dynamical measurements
constrain the total normalisation of the IMF and not its specific
form. Therefore, by themselves they cannot constrain the shape of the
IMF and reveal whether the change in the normalisation is due to a
change in the abundance of low mass stars or remnants of high mass
stars or any combination thereof. Independent constraints such as
those from stellar populations diagnostics or from galaxies at
different redshifts, probing different ranges in stellar mass are
necessary to break this degeneracy in the interpretation of the trend.

A final issue that we have not addressed in this paper is how much of
the trend shown in Figure~\ref{correlation1} is due to a varying IMF
and how much is it due to varying dark matter content. For the higher
velocity dispersion SLACS galaxies weak and strong lensing and
dynamical data allowed \citet{2010ApJ...721L.163A} to disentangle the
two effects, showing that the most likely explanation is that both the
dark matter content and the normalisation of the IMF increase with
galaxy mass \citep[see also][]{2009A&A...504..769C,
2010ApJ...721L...1T, 2011MNRAS.418.1557T, 2011arXiv1110.0833D,
2010MNRAS.402L..67G, 2011MNRAS.415..545T, 2011MNRAS.416..322D, 2012arXiv1201.2945T}.

Future papers of this series (Barnab\`e et al. 2012; Dutton et
al. 2012, in preparation) will combine the lensing information with
dynamical models of kinematic observations to construct two component
mass models and disentangle the luminous and dark components for the
SWELLS sample as well \citep[see ][]{2011MNRAS.417.1621D,
  2010MNRAS.401.1540T, 2010ApJ...719.1481V, 2011arXiv1110.2536S}. This
will allow us to differentiate trends in the stellar IMF from trends
in dark matter content over the entire range in velocity dispersion.

\section{Summary and conclusions}
\label{sec:sum}

In this paper we have presented complete multiband {\it HST} imaging
(including new images obtained as part of our cycle 18 {\it HST} program,
GO-12292) for the largest currently known sample of gravitational lens
systems where the deflector is an inclined discy galaxy. The SWELLS
sample consists to date of 20 secure lenses and 6 probable lenses.

For each secure lens and one probable lens, we have derived
gravitational lens models based on a single isothermal ellipsoid
deflector plus external shear. In addition we have used the
multicolour {\it HST} imaging to estimate the stellar mass of the bulge and
disc component of each deflector based on stellar population synthesis
models.  We have used the two measurements of stellar and total mass
within the critical curve to study the normalisation of the stellar
IMF for spiral lens galaxies.

Our main conclusions can be summarised as follows.

\begin{itemize}
\item The ratio between stellar mass and total mass within the
  critical curve correlates with the velocity dispersion of the
  deflector galaxy. By combining the SWELLS sample with the SLACS
  sample of more massive early-type galaxies we find that the trend
  extends between 160 and 350 $\kms$. This trend can be due to
  variations in the normalisation of the stellar IMF and/or in non
  baryonic dark matter content.

\item At the low velocity dispersion end (below 230 $\kms$) the
  Salpeter IMF predicts, unphysically, that the mass in stars is
  greater than the total mass.

\item Based on a rigorous probabilistic model, we find that the
  probability that the true stellar masses are lighter than implied by
  the Salpeter IMF is 98\% for galaxies with velocity dispersion below
  230 $\kms$, neglecting the contribution of cold gas. If the
  contribution of cold gas is taken into account the probability
  increases to greater than 99\%, and a 95\% upper limit on the
  stellar masses is 80\% of the mass implied by the Salpeter IMF.

\item The light IMF required for low mass spiral galaxies is at
  variance with the Salpeter or heavier IMF required for more massive
  early-type galaxies, consistent with a non universal stellar IMF.

\end{itemize}

\section*{Acknowledgements}

\input{acknowledgments.tex}

\appendix
\section{Comparison of Lens Models with Previous Models}
\begin{figure}
\vspace{0.5cm}
\begin{center}
\includegraphics[scale=0.46]{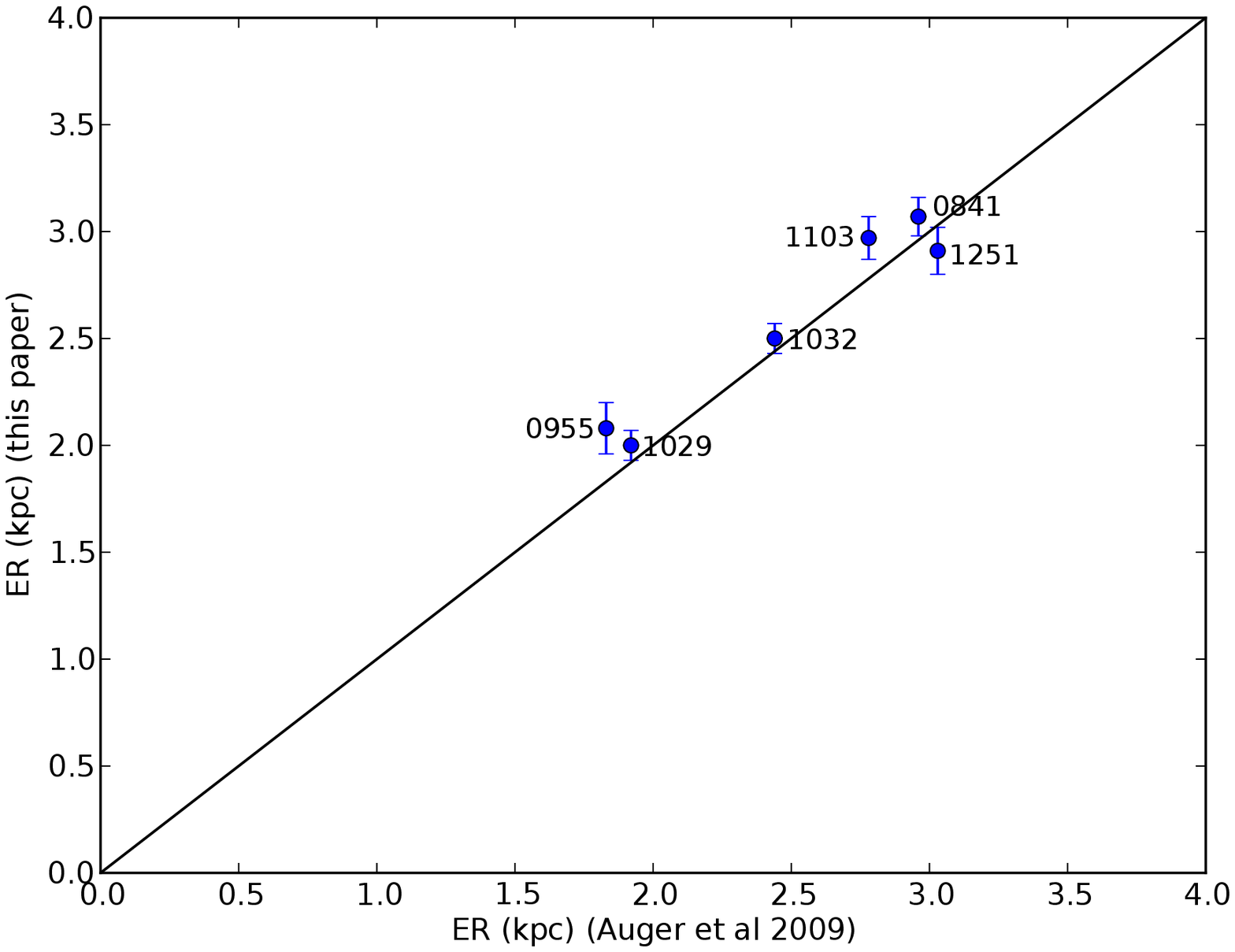}
\caption{A comparison of the inferred Einstein Radii of lenses
  included in the SWELLS sample and the SLACS sample. Error bars were
  not reported by \citet{2009ApJ...705.1099A}. The black line is the
  1-1 relation. The inferred Einstein Radii are consistent between the
  two studies.\label{duplicates}}
\end{center}
\end{figure}

The SWELLS sample contains six spiral lens galaxies that were
originally discovered and modelled in SLACS. As a sanity check, we
compare the inferred Einstein Radii of these lenses from the models
presented in this paper with those reported by
\citet{2009ApJ...705.1099A}. Since \citet{2009ApJ...705.1099A}
reported Einstein Radii in units of kiloparsecs, we convert our
Einstein Radii to kiloparsecs also.  The comparison is shown in
Figure~\ref{duplicates} and demonstrates agreement between the two
sets of models.




\label{lastpage}
\bsp

\end{document}

%% file: macros.tex
\newcommand {\apj} {ApJ}

\newcommand {\mnras} {MNRAS}

\newcommand {\etal} {et~al.~}
\def \spose#1{\hbox  to 0pt{#1\hss}}  
\newcommand {\lta} {\mathrel{\spose{\lower 3pt\hbox{$\sim$}}\raise  2.0pt\hbox{$<$}}}
\newcommand {\gta} {\mathrel{\spose{\lower  3pt\hbox{$\sim$}}\raise 2.0pt\hbox{$>$}}}

\newcommand {\ha}  {\ifmmode H\alpha \else H$\alpha $ \fi} 
 
\newcommand {\kms} {\ifmmode  \,\rm km\,s^{-1} \else $\,\rm km\,s^{-1}  $ \fi }
\newcommand {\kpc} {\ifmmode  {\rm kpc}  \else ${\rm  kpc}$ \fi  }  
\newcommand {\pc} {\ifmmode  {\rm pc}  \else ${\rm pc}$ \fi  }  
\newcommand {\Msun} {\ifmmode {\rm M_{\odot}} \else ${\rm M_{\odot}}$ \fi} 
\newcommand {\Zsun} {\ifmmode {\rm Z_{\odot}} \else ${\rm Z_{\odot}}$ \fi} 
\newcommand {\yr} {\ifmmode yr^{-1} \else $yr^{-1}$ \fi} 
\newcommand {\hMsun} {\ifmmode h^{-1}\,\rm M_{\odot} \else $h^{-1}\,\rm M_{\odot}$ \fi}

\def\zd{z_{\rm d}}
\def\zs{z_{\rm s}}

\def\q3{q_{3}}

\def\Mstar{M_{*}}

\def\SPASMOID{{\sc SPASMOID}\xspace}

\usepackage[usenames]{color}

%% file: addresses.tex
\def\uvic{Dept. of Physics and Astronomy, 
  University of Victoria, Victoria, BC, V8P 5C2, Canada}
\def\lick{UCO/Lick Observatory, Department of Astronomy and Astrophysics, 
  University of California, Santa Cruz, CA 95064, USA}
\def\ucsb{Dept. of Physics, University of California, 
  Santa Barbara, CA 93106, USA}
\def\kipac{Kavli Institute for Particle Astrophysics and Cosmology, 
 Stanford University, 452 Lomita Mall, Stanford, CA 94035, USA}
\def\utah{Department of Physics and Astronomy, University of Utah, 
  Salt Lake City, UT 84112, USA}
\def\kapteyn{Kapteyn Astronomical Institute, University of Groningen, 
  P.O.Box 800, 9700 AV Groningen, The Netherlands}
\def\oxford{Department of Physics, University of Oxford, 
  Keble Road, Oxford, OX1 3RH, UK}
\def\cambridge{Institute of Astronomy, University of Cambridge,
  Madingley Rd, Cambridge, CB3 0HA, UK}

\def\breweremail{\tt brewer@physics.ucsb.edu}

\def\packard{Packard Research Fellow}
\def\cita{CITA National Fellow}

%% file: catalogs/simple_table1.tex
J0820+4847  & 125.05363 &   48.79364 & 0.131 & 0.634 & 0.82$\pm$0.09 & 191.9$\pm$10.7 & 0.19$\pm$0.09 & 5.21$\pm$1.15 & 0.70$\pm$0.17& 1.18$\pm$0.28 & 3 & F450W \\ 
J0822+1828  & 125.614474&   18.482175& 0.1153& 0.8710& 0.83$\pm$0.03 & 185.1$\pm$4.0 & 0.75$\pm$0.08 & 4.36$\pm$0.28 & 0.54$\pm$0.17 & 1.03$\pm$0.39 & 4 & F435W \\ 
J0841+3824  & 130.37004 &   38.40381 & 0.116 & 0.657 & 1.46$\pm$0.05 & 251.2$\pm$4.4 & 0.70$\pm$0.08 & 14.1$\pm$0.99 & 0.51$\pm$0.11 & 0.92$\pm$0.18 & 2 & F814W \\ 
J0915+4211  & 138.81787 &   42.19800 & 0.078 & 0.790 & 0.98$\pm$0.02 & 195.7$\pm$2.2 & 0.64$\pm$0.08 & 4.03$\pm$0.18 & 0.81$\pm$0.18& 1.44$\pm$0.30& 3 & Kp \\ 
J0930+2855  & 142.572560&   28.916732& 0.3051& 0.4594& 1.05$\pm$0.02 & 342.1$\pm$2.8 & 0.81$\pm$0.08 & 40.7$\pm$1.33 & 0.28$\pm$0.09 & 0.51$\pm$0.19 & 4 & F814W \\  
J0955+0101  & 148.83217 &    1.02901 & 0.111 & 0.316 & 1.04$\pm$0.06 & 238.4$\pm$7.3 & 0.42$\pm$0.11 & 8.73$\pm$1.12 & 0.39$\pm$0.07 & 0.67$\pm$0.12 & 2 & F450W \\ 
J1021+2028  & 155.435084&   20.477728& 0.1208& 0.3507& 0.49$\pm$0.05 & 162.8$\pm$8.1 & 0.35$\pm$0.16 & 2.08$\pm$0.41 & 1.25$\pm$0.39 & 2.32$\pm$1.02 & 4 & F435W \\  
J1029+0420  & 157.34560 &    4.33384 & 0.104 & 0.615 & 1.05$\pm$0.03 & 211.7$\pm$3.3 & 0.72$\pm$0.15 & 6.54$\pm$0.42 & 0.68$\pm$0.13 & 1.19$\pm$0.26 & 1 & F814W \\ 
J1032+5322  & 158.14932 &   53.37636 & 0.133 & 0.329 & 1.06$\pm$0.03 & 251.0$\pm$3.3 & 0.58$\pm$0.14 & 11.25$\pm$0.56& 0.46$\pm$0.07 & 0.79$\pm$0.13 & 2 & F450W \\ 
J1037+3517  & 159.43764 &   35.29194 & 0.122 & 0.448 & ... & ... & ... & ... & ... & ... & 3 & ... \\
J1103+5322  & 165.78421 &   53.37450 & 0.158 & 0.735 & 1.08$\pm$0.03 & 222.8$\pm$3.1 & 0.49$\pm$0.10 & 10.65$\pm$0.65 & 0.61$\pm$0.10& 1.05$\pm$0.17 & 2 & F814W \\ 
J1111+2234  & 167.866760&   22.580729& 0.2223& 0.9887& 1.10$\pm$0.01 & 227.9$\pm$1.5 & 0.52$\pm$0.07 & 14.95$\pm$0.40 & 0.77$\pm$0.20 & 1.37$\pm$0.36 & 4 & F435W \\  
J1117+4704  & 169.39742 &   47.06873 & 0.169 & 0.405 & 0.77$\pm$0.02 & 217.5$\pm$2.5 & 0.59$\pm$0.14 & 7.83$\pm$2.16 & 0.61$\pm$0.16 & 1.10$\pm$0.32 & 3 & F450W \\ 
J1135+3720  & 173.77867 &   37.33997 & 0.162 & 0.402 & 0.71$\pm$0.07 & 206.3$\pm$10.3 & 0.22$\pm$0.10 & 6.14$\pm$1.22 & 0.60$\pm$0.16 & 1.11$\pm$0.32 & 3 & F435W \\  
J1203+2535  & 180.98470 &   25.59697 & 0.101 & 0.856 & 0.87$\pm$0.06 & 187.2$\pm$5.9 & 0.80$\pm$0.10 & 4.17$\pm$0.50 & 0.68$\pm$0.14 & 1.15$\pm$0.20 & 3 & F606W \\  
J1251$-$0208& 192.89877 & $-$2.13477 & 0.224 & 0.784 & 0.82$\pm$0.03 & 203.0$\pm$2.6 & 0.74$\pm$0.11 & 8.89$\pm$0.46 & 0.47$\pm$0.08 & 0.86$\pm$0.11 & 1 & F606W \\  
J1313+0506  & 198.36127 &    5.11589 & 0.144 & 0.338 & 0.53$\pm$0.23 & 172.2$\pm$22.1 & 0.79$\pm$0.16 & 2.91$\pm$1.29 & 1.11$\pm$0.54 & 1.86$\pm$0.95& 2 & F606W \\  
J1331+3638  & 202.91800 &   36.46999 & 0.113 & 0.254 & 0.96$\pm$0.04 & 248.1$\pm$4.4 & 0.67$\pm$0.09 & 8.86$\pm$0.61 & 0.38$\pm$0.08 & 0.71$\pm$0.16& 3 & F450W \\ 
J1422+4134  & 215.609046&   41.576612& 0.1011& 3.01  & 1.18$\pm$0.05 & 208.5$\pm$4.3 & 0.73$\pm$0.22 & 7.01$\pm$0.58 & 0.13$\pm$0.05 & 0.23$\pm$0.08 & 4 & F814W \\ 
J1703+2451  & 255.92278 &   24.86111 & 0.063 & 0.637 & 0.93$\pm$0.05 & 189.7$\pm$2.2 & 0.42$\pm$0.08 & 2.91$\pm$0.11 & 0.45$\pm$0.06 & 0.84$\pm$0.10& 3 & Kp \\  
J2141$-$0001& 325.47781 & $-$0.02008 & 0.138 & 0.713 & 0.88$\pm$0.09 & 197.0$\pm$11.0 & 0.48$\pm$0.21 & 6.07$\pm$1.34 & 0.67$\pm$0.15 & 1.14$\pm$0.26 & 2 & F814W \\

%% file: catalogs/simple_table2.tex
J0007+0053  &   1.886461&    0.889701& 0.2752& 0.9119& X &  0.54 & 4 & F435W/F814W \\
J0028$-$0014&   7.220414& $-$0.243334& 0.1115& 0.3103& C &  0.53 & 4 & F160W\\ 
J0149$-$0010&  27.401391& $-$0.169114& 0.2437& 0.8263& C &  0.57 & 4 & F435W/F814W \\
J0329$-$0027&  52.465504& $-$0.460612& 0.2429& 0.5117& B &  0.53 & 4 & F435W/F814W \\
J0745+3400  & 116.428559&   34.007752& 0.0647& 0.2953& X &  0.36 & 4 & F160W\\
J0821+1025  & 125.48858 &   10.43226 & 0.0942& 0.6568& C &  ...  & 3 & F435W/F814W \\ 
J0822+1828  & 125.614474&   18.482175& 0.1153& 0.8710& A &  0.46 & 4 & F435W/F814W \\
J0825+2109  & 126.320762&   21.161022& 0.1012& 0.3956& X &  0.26 & 4 & F435W/F814W \\
J0908+2730  & 137.234813&   27.513829& 0.0219& 0.2831& X &  0.15 & 4 & F160W\\
J0915+4211  & 138.81787 &   42.19800 & 0.078 & 0.790 & A &  0.59 & 3 & F435W/F814W \\ 
J0930+2855  & 142.572560&   28.916732& 0.3051& 0.4594& A &  0.55 & 4 & F435W/F814W \\
J1000+2835  & 150.011213&   28.595699& 0.0885& 0.8757& X &  0.41 & 4 & F435W/F814W \\
J1021+2028  & 155.435084&   20.477728& 0.1208& 0.3507& A &  0.39  & 4 & F435W/F814W \\
J1111+2234  & 167.866760&   22.580729& 0.2223& 0.9887& A &  0.58 & 4 & F435W/F814W \\
J1135+3720  & 173.77867 &   37.33997 & 0.162 & 0.402 & A &  0.32 & 3 & F435W/F814W \\
J1249+0225  & 192.467683&    2.417972& 0.0965& 0.3517& C &  0.43 & 4 & F435W/F814W \\
J1403+1530  & 210.799017&   15.511061& 0.0976& 1.0858& C &  ...  & 4 & F435W/F814W/F160W\\
J1422+4134  & 215.609046&   41.576612& 0.1011& 3.01  & B &  0.20 & 4 & F435W/F814W \\
J1629+4708  & 247.291334&   47.145309& 0.1282& 0.3739& B &  0.41 & 4 & F435W/F814W \\
J1633+1341  & 248.414833&   13.687109& 0.1171& 0.2436& C &  0.31 & 4 & F435W/F814W \\
J2333$-$1042& 353.272047&$-$10.703289& 0.1445& 0.1328& B &  0.37 & 4 & F435W/F814W \\

%% file: catalogs/simple_table3.tex
SDSSJ0007$+$0053 &  21.79  &    19.12  &  \nodata  &  \nodata & 21.79  &    19.72  &  \nodata  &  \nodata & 0.45 & 0.68 & 1.40 & 0.39\\ 
SDSSJ0149$-$0010 &  22.96  &    20.09  &  \nodata  &  \nodata & 22.93  &    20.56  &  \nodata  &  \nodata & 0.21 & 0.56 & 0.39 & 0.48\\
SDSSJ0329$-$0027 &  22.68  &    19.63  &  \nodata  &  \nodata & 21.91  &    19.52  &  \nodata  &  \nodata & 0.16 & 0.85 & 0.90 & 0.39\\ 
SDSSJ0821$+$1025 &  20.69  &    17.57  &  \nodata  &  \nodata & 18.75  &    16.81  &  \nodata  &  \nodata &  1.28 & 0.52 & 2.59 & 0.15\\   
SDSSJ0822$+$1828 &  19.86  &    17.50  &  \nodata  &  \nodata & 20.51  &    18.40  &  \nodata  &  \nodata &  0.65 & 0.66 & 1.09 & 0.24\\   
SDSSJ0825$+$2109 &  48.84  &    20.35  &  \nodata  &  \nodata & 18.80  &    16.99  &  \nodata  &  \nodata &  0.32 & 0.79 & 1.80 & 0.24\\   
SDSSJ0915$+$4211 &  19.02  &    16.82  &  \nodata  &    15.63 & 19.55  &    17.68  &  \nodata  &    16.73  &  1.05 & 0.89 & 1.40 & 0.25\\  
SDSSJ0930$+$2855 &  21.53  &    18.28  &  \nodata  &  \nodata & 21.97  &    18.95  &  \nodata  &  \nodata &  0.62 & 0.53 & 2.36 & 0.75\\   
SDSSJ1000$+$2835 &  19.56  &    17.32  &  \nodata  &  \nodata & 19.61  &    17.71  &  \nodata  &  \nodata &  0.58 & 0.79 & 1.43 & 0.20\\   
SDSSJ1021$+$2028 &  19.75  &    17.42  &  \nodata  &  \nodata & 20.82  &    19.13  &  \nodata  &  \nodata &  0.40 & 0.44 & 0.70 & 0.17\\  
SDSSJ1111$+$2234 &  20.33  &    17.62  &  \nodata  &  \nodata & 19.94  &    17.84  &  \nodata  &  \nodata &  0.71 & 0.67 & 1.38 & 0.46\\  
SDSSJ1135$+$3720 &  21.37  &    18.45  &  \nodata  &    17.19 & 19.91  &    17.54  &  \nodata  &    16.45 &  0.56 & 0.67 & 1.41 & 0.26 \\   
SDSSJ1249$+$0225 &  20.31  &    18.08  &  \nodata  &  \nodata & 19.78  &    17.84  &  \nodata  &  \nodata &  0.37 & 0.73 & 1.06 & 0.32\\  
SDSSJ1403$+$1530 &  20.95  &    18.16  &    16.88  &  \nodata & 18.98  &    17.02  &    16.34  &  \nodata &  0.60 & 0.62 & 1.97 & 0.36\\  
SDSSJ1422$+$4134 &  48.24  &    19.69  &  \nodata  &  \nodata & 18.64  &    16.78  &  \nodata  &  \nodata &  0.54 & 0.57 & 2.06 & 0.16\\  
SDSSJ1629$+$4708 &  20.17  &    17.80  &  \nodata  &  \nodata & 19.64  &    17.56  &  \nodata  &  \nodata &  0.39 & 0.71 & 1.16 & 0.25\\  
SDSSJ1633$+$1341 &  21.90  &    19.29  &  \nodata  &  \nodata & 19.69  &    17.38  &  \nodata  &  \nodata &  0.91 & 0.15 & 2.07 & 0.26\\  
SDSSJ2333$-$1042 &  20.40  &    16.97  &  \nodata  &  \nodata & 19.48  &    17.13  &  \nodata  &  \nodata &  1.23 & 0.65 & 1.32 & 0.18\\  

%% file: catalogs/simple_table4.tex
SDSSJ0007$+$0053 &  $10.63 \pm 0.14$ & $10.26 \pm 0.11$ &  $10.84 \pm 0.18$ & $10.43 \pm 0.13$  \\ 
SDSSJ0149$-$0010 &  $10.21 \pm 0.16$ & $ 9.90 \pm 0.14$ &  $10.57 \pm 0.19$ & $10.21 \pm 0.13$  \\
SDSSJ0329$-$0027 &  $10.44 \pm 0.14$ & $10.26 \pm 0.17$ &  $10.72 \pm 0.15$ & $10.57 \pm 0.15$  \\ 
SDSSJ0821$+$1025 &  $10.40 \pm 0.11$ & $10.60 \pm 0.13$ &  $10.64 \pm 0.11$ & $10.78 \pm 0.16$  \\   
SDSSJ0822$+$1828 &  $10.54 \pm 0.13$ & $10.09 \pm 0.11$ &  $10.79 \pm 0.16$ & $10.34 \pm 0.16$  \\   
SDSSJ0825$+$2109 &      ...          & $10.58 \pm 0.17$ &     ...           & $10.81 \pm 0.16$  \\   
SDSSJ0915$+$4211 &  $10.60 \pm 0.11$ & $10.17 \pm 0.10$ &  $10.83 \pm 0.09$ & $10.43 \pm 0.09$ \\  
SDSSJ0930$+$2855 &  $11.18 \pm 0.15$ & $10.85 \pm 0.15$ &  $11.42 \pm 0.17$ & $11.11 \pm 0.17$  \\   
SDSSJ1000$+$2835 &  $10.40 \pm 0.14$ & $10.18 \pm 0.14$ &  $10.57 \pm 0.17$ & $10.34 \pm 0.14$  \\   
SDSSJ1021$+$2028 &  $10.62 \pm 0.13$ & $ 9.74 \pm 0.14$ &  $10.87 \pm 0.18$ & $ 9.98 \pm 0.12$  \\  
SDSSJ1111$+$2234 &  $11.11 \pm 0.13$ & $10.88 \pm 0.15$ &  $11.37 \pm 0.14$ & $11.10 \pm 0.15$  \\  
SDSSJ1135$+$3720 &  $10.62 \pm 0.08$ & $10.91 \pm 0.10$ &  $10.85 \pm 0.09$ & $11.09 \pm 0.07$ \\   
SDSSJ1249$+$0225 &  $10.15 \pm 0.13$ & $10.16 \pm 0.13$ &  $10.32 \pm 0.16$ & $10.50 \pm 0.15$  \\  
SDSSJ1403$+$1530 &  $10.33 \pm 0.10$ & $10.51 \pm 0.12$ &  $10.58 \pm 0.12$ & $10.76 \pm 0.10$  \\  
SDSSJ1422$+$4134 &      ...          & $10.56 \pm 0.17$ &      ...          & $10.84 \pm 0.15$  \\  
SDSSJ1629$+$4708 &  $10.52 \pm 0.12$ & $10.49 \pm 0.15$ &  $10.73 \pm 0.16$ & $10.78 \pm 0.12$  \\  
SDSSJ1633$+$1341 &  $ 9.88 \pm 0.14$ & $10.67 \pm 0.15$ &  $10.11 \pm 0.13$ & $10.84 \pm 0.20$  \\  
SDSSJ2333$-$1042 &  $11.08 \pm 0.13$ & $10.86 \pm 0.14$ &  $11.29 \pm 0.12$ & $11.07 \pm 0.15$  \\  

%% file: catalogs/simple_table5.tex
SDSSJ0007$+$0053   & 2.37$\pm$0.70  & 2.02$\pm$0.50   & 1.37$\pm$0.33  & 1.35$\pm$0.28    &  3.98$\pm$1.54  & 3.38$\pm$1.03   & 2.05$\pm$0.55   & 2.06$\pm$0.45 \\
SDSSJ0149$-$0010   & 3.36$\pm$1.07  & 2.65$\pm$0.71   & 2.01$\pm$0.58  & 1.76$\pm$0.40    &  7.40$\pm$3.00  & 5.58$\pm$1.82   & 4.02$\pm$1.12   & 3.55$\pm$0.78 \\
SDSSJ0329$-$0027   & 3.38$\pm$0.95  & 2.66$\pm$0.61   & 1.65$\pm$0.65  & 1.55$\pm$0.51    &  6.62$\pm$2.06  & 5.14$\pm$1.28   & 3.28$\pm$1.02   & 2.94$\pm$0.74 \\
SDSSJ0821$+$1025   & 6.48$\pm$1.73  & 4.52$\pm$1.05   & 2.92$\pm$0.91  & 2.44$\pm$0.64    & 11.66$\pm$2.79  & 8.20$\pm$1.67   & 4.49$\pm$1.78   & 3.82$\pm$1.18 \\
SDSSJ0822$+$1828   & 3.45$\pm$1.14  & 2.75$\pm$0.73   & 2.48$\pm$0.68  & 2.19$\pm$0.47    &  6.30$\pm$2.28  & 5.04$\pm$1.45   & 4.49$\pm$1.70   & 3.80$\pm$1.14 \\
SDSSJ0825$+$2109   & \nodata	    & \nodata         & 2.43$\pm$1.01  & 2.11$\pm$0.67    &  \nodata        & \nodata         & 4.16$\pm$1.54   & 3.62$\pm$1.05 \\
SDSSJ0915$+$4211   & 5.16$\pm$1.40  & 3.82$\pm$0.87   & 3.54$\pm$0.97  & 2.84$\pm$0.63    &  8.45$\pm$2.11  & 6.39$\pm$1.33   & 6.41$\pm$1.58   & 5.10$\pm$1.05 \\
SDSSJ0930$+$2855   & 3.70$\pm$1.03  & 2.84$\pm$0.65   & 3.01$\pm$0.88  & 2.37$\pm$0.55    &  6.55$\pm$2.18  & 4.98$\pm$1.35   & 5.54$\pm$1.83   & 4.38$\pm$1.17 \\
SDSSJ1000$+$2835   & 4.12$\pm$1.33  & 3.16$\pm$0.82   & 2.98$\pm$0.99  & 2.47$\pm$0.67    &  6.33$\pm$2.69  & 4.97$\pm$1.69   & 4.36$\pm$1.49   & 3.74$\pm$1.00 \\
SDSSJ1021$+$2028   & 3.65$\pm$1.04  & 2.85$\pm$0.64   & 1.69$\pm$0.61  & 1.60$\pm$0.45    &  6.59$\pm$2.66  & 5.19$\pm$1.72   & 2.91$\pm$0.73   & 2.77$\pm$0.59 \\
SDSSJ1111$+$2234   & 3.23$\pm$0.85  & 2.56$\pm$0.54   & 1.87$\pm$0.60  & 1.71$\pm$0.45    &  5.97$\pm$2.00  & 4.71$\pm$1.27   & 3.08$\pm$0.88   & 2.77$\pm$0.62 \\
SDSSJ1135$+$3720   & 5.63$\pm$1.16  & 4.05$\pm$0.73   & 3.72$\pm$0.91  & 2.92$\pm$0.62    &  9.51$\pm$2.24  & 6.86$\pm$1.41   & 5.46$\pm$1.10   & 4.41$\pm$0.73 \\
SDSSJ1249$+$0225   & 3.74$\pm$1.11  & 2.99$\pm$0.70   & 2.53$\pm$0.85  & 2.20$\pm$0.57    &  5.49$\pm$2.08  & 4.47$\pm$1.34   & 5.65$\pm$1.92   & 4.60$\pm$1.25 \\
SDSSJ1403$+$1530   & 7.25$\pm$1.85  & 5.00$\pm$1.12   & 2.66$\pm$0.72  & 2.25$\pm$0.48    & 13.20$\pm$4.08  & 9.06$\pm$2.43   & 4.65$\pm$1.18   & 3.90$\pm$0.83 \\
SDSSJ1422$+$4134   & \nodata        & \nodata         & 2.22$\pm$0.92  & 1.96$\pm$0.63    & \nodata         & \nodata         & 4.28$\pm$1.55   & 3.65$\pm$1.07 \\
SDSSJ1629$+$4708   & 3.58$\pm$1.01  & 2.83$\pm$0.64   & 2.42$\pm$0.81  & 2.06$\pm$0.56    &  5.97$\pm$2.03  & 4.72$\pm$1.33   & 4.57$\pm$1.22   & 3.88$\pm$0.85 \\
SDSSJ1633$+$1341   & 4.17$\pm$1.24  & 3.18$\pm$0.76   & 3.83$\pm$1.21  & 2.97$\pm$0.76    &  7.14$\pm$2.10  & 5.48$\pm$1.30   & 5.80$\pm$2.41   & 4.61$\pm$1.55 \\
SDSSJ2333$-$1042   & 7.13$\pm$1.99  & 4.92$\pm$1.21   & 3.11$\pm$0.95  & 2.51$\pm$0.61    & 11.80$\pm$3.38  & 8.20$\pm$2.03   & 5.13$\pm$1.73   & 4.21$\pm$1.14 \\

%% file: acknowledgments.tex
AAD acknowledges financial support from a CITA National Fellowship,
from the National Science Foundation Science and Technology Center
CfAO, managed by UC Santa Cruz under cooperative agreement
No. AST-9876783. AAD and DCK were partially supported by NSF grant AST
08-08133, and by HST grants AR-10664.01-A, HST AR-10965.02-A, and HST
GO-11206.02-A.
PJM was given support by the TABASGO and Kavli foundations, and the Royal 
Society, in the form of research fellowships.
TT acknowledges support from the NSF through CAREER award NSF-0642621,
and from the Packard Foundation through a Packard Research Fellowship.
LVEK acknowledges the support by an NWO-VIDI programme subsidy
(programme number 639.042.505).
This research is supported by NASA through Hubble Space Telescope
programs GO-10587, GO-10886, GO-10174, 10494, 10798, 11202, 11978,
12292 and in part by the National Science Foundation under Grant
No. PHY99-07949. and is based on observations made with the NASA/ESA
Hubble Space Telescope and obtained at the Space Telescope Science
Institute, which is operated by the Association of Universities for
Research in Astronomy, Inc., under NASA contract NAS 5-26555, and at
the W.M. Keck Observatory, which is operated as a scientific
partnership among the California Institute of Technology, the
University of California and the National Aeronautics and Space
Administration. The Observatory was made possible by the generous
financial support of the W.M. Keck Foundation. The authors wish to
recognize and acknowledge the very significant cultural role and
reverence that the summit of Mauna Kea has always had within the
indigenous Hawaiian community.  We are most fortunate to have the
opportunity to conduct observations from this mountain.
Funding for the SDSS and SDSS-II was provided by the Alfred P. Sloan
Foundation, the Participating Institutions, the National Science
Foundation, the U.S. Department of Energy, the National Aeronautics
and Space Administration, the Japanese Monbukagakusho, the Max Planck
Society, and the Higher Education Funding Council for England. The
SDSS was managed by the Astrophysical Research Consortium for the
Participating Institutions. The SDSS Web Site is http://www.sdss.org/.